\documentclass[preprint2,times,tighten,twocolappendix]{aastex6}

\usepackage{color}
\usepackage{enumitem}
\usepackage{hyperref}
\usepackage{verbatim}
\usepackage{amsmath,amstext,mathrsfs}
\usepackage[all]{hypcap} 
\usepackage{afterpage}
\usepackage{marginnote}

\begin{document}
\title{Synchrotron intensity gradients as tracers of interstellar magnetic field}

\author{A. Lazarian\altaffilmark{1}, Ka Ho Yuen\altaffilmark{1,2}, Hyeseung Lee\altaffilmark{1,3}, J. Cho\altaffilmark{1,3}}
%\email{lazarian@astro.wisc.edu, }
\altaffiltext{1}{Department of Astronomy, University of Wisconsin-Madison}
\altaffiltext{2}{Department of Physics, The Chinese University of Hong Kong}
\altaffiltext{3}{Department of Physics, Chugnam University, Korea}
\begin{abstract}
On the basis of the modern understanding of MHD turbulence, we propose a new way of using synchrotron radiation, namely using synchrotron intensity gradients for tracing astrophysical magnetic fields.  We successfully test the new technique using synthetic data obtained with the 3D MHD simulations and provide the demonstration of the practical utility of the technique by comparing the directions of magnetic field that are obtained with PLANCK synchrotron intensity datas to the directions obtained with PLANCK synchrotron polarization data. We demonstrate that the synchrotron intensity gradients (SIGs) can reliably trace magnetic field in the presence of noise and can provide detailed maps of magnetic-field directions.  We also show that the SIGs are relatively robust for tracing magnetic fields while the low spacial frequencies of the synchrotron image are removed. This makes the SIGs applicable to tracing of magnetic fields using interferometric data with single dish measurement absent. We discuss the synergy of using the SIGs together with synchrotron polarization in order to find the actual direction of the magnetic field and quantify the effects of Faraday rotation as well as with other ways of studying astrophysical magnetic fields. We test our method in the presence of noise and the resolution effects. We stress the complementary nature of the studies using the SIG technique and those employing the recently-introduced velocity gradient techniques that traces the magnetic fields using spectroscopic data. 
\end{abstract}

\keywords{ISM: general, structure --- MHD --- radio continuum: ISM --- turbulence}
\section{Introduction}

This paper provides a description of a new technique for studying magnetic fields using gradients of synchrotron intensity. Gradients of synchrotron polarization have been successfully used before (see \citealt{Gaensler2011Low-Mach-numberGradients, Burkhart2012PropertiesMaps}). However, in this Letter we explore theoretically and numerically a more simple measure, namely, synchrotron intensity gradients (SIGs) and evaluate its utility for observational study of magnetic fields and accounting for the foreground contamination induced by the interstellar media within the CMB polarization studies. 

Galactic and Extragalactic synchrotron emission arises from relativistic electrons moving in astrophysical magnetic field (see \citealt{1981MoIzNRG}). In terms of CMB and high redshift HI studies, the most important is galactic synchrotron emission. However, diffuse synchrotron emission is also a major emission arising from the interstellar medium (ISM), the intracluster medium (ICM), as well as in the lobes of radio galaxies (e.g.  \citealt{2006ApJS..167..230H,2006AJ....131.2900C,2007A&A...471L..21S,2008MNRAS.391..521L}). In fact, synchrotron emission provides the largest range of scales for studying magnetic fields. 

Astrophysical magnetic fields are turbulent as observations testify that turbulence is ubiquitous in astrophysics
\citep{Armstrong1995ElectronMedium, 2009SSRv..143..357L, Chepurnov2010ExtendingData}. As relativistic electrons are present in most cases, the turbulence results in synchrotron fluctuations, which may provide detailed information about magnetic fields at different scales, but, at the same time, interfere with the measurements of CMB and high redshift HI. The latter has recently become a topic of intensive discussion (see \citealt{2008PhRvL.100p1301L,2009ApJLiu,2014MNRAS.440..298F}).

The statistics of synchrotron intensity has been studied recently in \citeauthor{LP12} (\citeyear{LP12}, hereafter LP12), where it was shown how fluctuations of synchrotron intensity can be related to the fluctuations of magnetic field for an arbitrary index of cosmic rays spectrum. There it was shown that the turbulence imprints its anisotropy on synchrotron intensity and this provides a way of determining the direction of the mean magnetic field using synchrotron intensities only. The current paper explores whether, on the basis of our present-day understanding of the nature of MHD turbulence, {\it synchrotron intensities} can provide more detailed information about magnetic fields.

In what follows \S \ref{sec:2} we discuss the theoretical motivation of this work routed in the modern theory of the  of MHD turbulence,
the properties of synchrotron intensity gradients (SIGs), their calculation as well as the influence of noise and sonic Mach number are discussed in \S \ref{sec:3}. We illustrate our method on PLANCK synchrotron data at \S \ref{sec:4} The comparison of the SIGs technique with the technique based on the anisotropy of the correlation functions of intensity is presented in \S \ref{sec:5}, the synergy with other techniques of magnetic field studies is outlined in \S \ref{sec:6}. We present our summary in \S \ref{sec:7}.

\section{Theoretical considerations}
\label{sec:2}

\subsection{MHD turbulence and magnetic field gradients}

Dealing with synchrotron emitting media, we deal with the non-relativistic thermal magnetized plasma and relativistic electrons. The turbulence in magnetized relativistic and non-relativistic fluids are different (see \citealt{2016ApJ...831L..11T}). However, following the accepted approach, we consider turbulence in magnetized fluid separately from the fluid of cosmic rays. In other words, we consider that relativistic electrons just illuminate the structure of magnetic field that is created by non-relativistic MHD turbulence. This approach has its limitations (see \citealt{LB06,YL11}) but for our further discussion this is not critically important.

While the original studies of Alfvenic turbulence done by \cite{I64} and \cite{K65} were based a hypothetical model of isotropic MHD turbulence, the 
later studies (see \citealt{1981PhFl...24..825M, 1983PhRvL..51.1484M, 1983JPlPh..29..525S,1984ApJ...285..109H}) uncovered the anisotropic nature of the MHD cascade.The modern theory of MHD turbulence arises from the prophetic work by \citealt{GoldreichP.Sridhar1995GS95IITurbulence}, henceforth GS95).  Further theoretical and numerical studies (\citealt{Lazarian1999ReconnectionField}, henceforth LV99, \citealt{Cho2000TheTurbulence,Maron2000SimulationsTurbulence,Lithwick2001CompressiblePlasmas,Cho2001SimulationsMedium,Cho2002CompressiblePlasmasb,Cho2003CompressibleImplicationsb,Kowal2010VelocityScalingsb}, see \citealt{Brandenburg2013AstrophysicalTurbulence} for a a review) extended the theory and augmented it with new concepts. Our theoretical motivation for the present work is based on the modern understanding of the nature of MHD turbulence that we briefly summarize below.

The GS95 theory treats the Alfvenic incompressible turbulence. The numerical simulations in \cite{Cho2002CompressiblePlasmasb} testify that for non-relativistic MHD turbulence the energy exchange between different types of fundamental modes is the effect that can be frequently neglected.\footnote{This is in contrast with the relativistic MHD turbulence where the coupling between fast and Alfvenic fundamental modes were shown to be significant \citep{2016ApJ...831L..11T}.} Therefore, in non-relativistic compressible astrophysical media one can consider three distinct cascades, namely, the cascade of Alfven, slow and fast modes.\footnote{We use the word "modes" rather than "waves", as the properties of the magnetic perturbations can be very different from those of waves. For instance, Alfven modes in GS95 turbulence are not oscillatory and after one period undergo cascading.} Therefore the GS95 treatment is applicable to describing Alfvenic modes also in compressible fluids.

Alfven modes initially evolve by increasing the perpendicular wavenumber in the subAlfvenic regime, i.e. for the injection velocity $v_L$ being less than the Alfven velocity $V_A$,  while the parallel wavenumber stays the same. This is the regime of weak turbulence with the spectrum $E(k)\sim k^{-2}$ (see LV99, \citealt{Gal2005}). This is not yet the regime of GS95 turbulence, but, nevertheless, the increase of the perpendicular wave number means the modes get more and more perpendicular to the magnetic field. In Alfvenic turbulence, the magnetic field and velocity are symmetric and therefore the aforementioned situation means that both the gradients of magnetic field and velocity is getting aligned perpendicular to the direction of the magnetic field. 

Weak Alfvenic turbulence can be viewed as the interaction of wave packets with a fraction of the energy of cascading as a result of such an interaction. As the perpendicular wavenumber increases, this fraction gets larger and eventually becomes $\sim 1$ (see \citealt{Cho2003CompressibleImplicationsb,L16}). This is the maximal fraction of energy that can be transferred during the wavepacket interaction. However, the equations dictate the necessity of further increase of perpendicular wavenumber as the result of the interaction of oppositely moving wavepackets. This can only be accomplished through the simultaneous increase of the parallel wavenumber. This happens at the transition scale $l_{trans}\approx L M_A^2$, where $L$ is the turbulence injection scale and $M_A=v_L/V_A$ is the Alfven Mach number (see LV99, \citealt{Lazarian2006}). This is the stage of the transfer of the strong or GS95 regime of turbulence. At this stage, the so-called critical balance condition should be satisfied, which states that the time of the interaction of the oppositely moving wavepackets
$l_{\|}/V_A$, where $l_{\|}$ is the parallel to magnetic field scale of the wavepacket is equal to the perpendicular shearing time of the wavepacket $l_{\bot}/v_l$, where $l_{\bot}$ is the perpendicular scale of the wavepacket and $v_l$ is the turbulent velocity associated with this scale. For the subAlfvenic turbulence this is how the cascade proceeds in the strong regime with the wavepackets getting more and more elongated according to (see LV99):
\begin{equation}
l_{\|}\approx L \left(\frac{l_{\bot}}{L}\right)^{2/3} M_A^{-4/3},
\label{lpar}
\end{equation}
which testifies that for $l_{\bot}\ll L$ the parallel scale of the wavepackets gets much larger than the perpendicular scale. This means that
the wavepackets/eddies get more and more elongated as the perpendicular scale $l_{\bot}$ decreases.  As a result, both the velocity and magnetic field gradients get more and more aligned perpendicular to the magnetic field direction as $l_{\bot}$ decreases. In fact, the increase of the disparity of the parallel and perpendicular scales continues until the energy reaches the dissipation scale. 

The magnetic field direction is changing in the turbulent flow. Therefore the important question that arises is what the direction of the magnetic field should be used in the arguments above. In other words, it is important to understand how the parallel and perpendicular directions to measure ${\|}$ and ${\bot}$ are defined. 
Most of the earlier MHD turbulence work assumed the perturbative approach and thus the mean field direction was used. This was also an implicit assumption in GS95 study. However, in the studies that followed the groundbreaking GS95 paper (namely, LV99, \citealt{Cho2000TheTurbulence, Maron2000SimulationsTurbulence,Cho2002CompressiblePlasmasb}) it was shown that it is not correct to measure the directions in respect to the mean magnetic field. Instead, one should use the {\it local} magnetic field that surrounds the wavepacket  $[l_{\|}, l_{\bot}]$. 

The importance of using of local system of reference is the most evident if arguments related to the fast turbulent reconnection are employed (LV99). Indeed, there it was shown in LV99 that the magnetic reconnection happens within a turnover time of an eddy and therefore the motions of fluid perpendicular to the magnetic field lines are similar to hydrodynamic eddies. As magnetic field lines reconnect fast, the mixing motion of perpendicular to the local direction of magnetic field does not create magnetic tension. As a result, the formation of such eddies provides the path of least resistance compared to any other motions involving magnetic field line bending. Naturally, the turbulent energy is channeled along this path of the least resistance. The hydrodynamic-type cascade of energy associated of these eddies is $\sim v_{l}^2/t_{cas}=const$, with the cascading time given by the eddy turnover time $l_{\bot}/v_{l}$. From what we said above, it is evident that $l_{\bot}$ has to be measured perpendicular to the magnetic field direction at the location of the eddy, rather than to the direction of the mean magnetic field. Incidentally, this hydrodynamic-type cascading that we describe provides the Kolmogorov scaling for the perpendicular turbulent motions, i.e. from $v_{\bot}\sim l_{\bot}^{1/3}$, that is the GS95 prediction. As the eddies rotate around the local magnetic field direction they flop sending the Alfven waves along the magnetic field. The corresponding Alfven wave period is equal to the eddy turnover time, i.e. $l_{\|}/V_A\approx l_{\bot}/v_{l}$. The latter coincides with the "critical balance" condition in GS95. Combining this with the Kolmogorov scaling of the perpendicular motions, one can get the $l_{\l}\sim l_{\bot}^{2/3}$ (see Eq. (\ref{lpar}). We would like to stress that this derivation dictates that $l_{\|}$ is aligned with the magnetic field of the eddy and $l_{\bot}$ is the size of the eddy perpendicular to the {\it local} magnetic field. The corresponding numerical studies (\citealt{Cho2000TheTurbulence, Maron2000SimulationsTurbulence,Cho2002CompressiblePlasmasb}) show that the aforementioned relations between $l_{\|}$ and $l_{\bot}$ are correct only in the {\it local} system of reference given by local magnetic field that the eddy interacts with.\footnote{The GS95 relation between $l_{\bot}$ and $l_{\|}$ isn{\it not} valid in the system of reference related to the mean magnetic field. Observations in most cases sample turbulence along the line of sight over distances much larger than the scale of the sampled eddy. In this situation the measurements take place in respect to the mean magnetic field (see \citealt{LP12}) and this prevents the observational testing the GS95 anisotropy.}
 
The anisotropy of turbulence given by the aspect ratio of the eddies (see Eq. (\ref{lpar})) is increasing with the decrease of the scale. Therefore the smaller the eddy the better it traces the local direction of the magnetic field. At the same time, one can estimate the velocity gradients that scale as $v_l/l_{\bot}\sim l_{\bot}^{-2/3}$. The latter relation means that the largest gradients correspond to the smallest eddies. Thus 3D gradients in Alfvenic turbulence should be dominated by the gradients of the smallest eddies and therefore the measured gradients should be perpendicular to the local direction of the magnetic field as it is traced by the smallest resolved eddies. Magnetic field and velocity enter in a symmetric way in Alfvenic turbulence and therefore the gradients of turbulent magnetic field should have the same property as the gradients of velocity. As gradients are linear operation then if we have a quantity that is an integral of magnetic fluctuations along the line of sight, as this is the case of synchrotron fluctuations, the gradient and integral operation can be interchanged and thus the observed 2D measure can be presented as an integral of magnetic field gradients. This shows that the observed quantities, which could be the integrated along the line of sight components of fluctuating velocities as in our papers on tracing magnetic field with velocity gradients \cite{GL17, YL17,YL17b,LY17}. In our present paper that suggests the way of magnetic field tracing with synchrotron, the gradients are dominated by the smallest eddies, that are most aligned with magnetic field. 

Naturally, Eq. (\ref{lpar}) provides only the most probable relation between the parallel and perpendicular scales. The distribution function relating $l_{\bot}$ and $l_{\|}$ was obtained in \cite{Cho2002CompressiblePlasmasb} on the basis of numerical simulations. Its analysis, however, shows that the uncertainty in the gradient direction is of the order of $l_{\bot}/l_{\|}\sim (l_{\bot}/L)^{1/3}$ indicating that the smaller the resolved eddies, the better the gradients trace magnetic fields. In other words, the most probable directions of the Alfven wavevectors are limited by the GS95 cone given by Eq. (\ref{lpar}), while the probability of the vectors to be beyond this cone is exponentially suppressed.\footnote{\cite{Cho2002CompressiblePlasmasb} argued that the actual distribution can be best represented by the Castaing function \cite{1990PhyD...46..177C} which is smooth near zero but looks like an exponential over a broad range.}
%Castaing, B., Gagne, Y., & Hop�nger, E. 1990, Physica D, 46, 177

We note that all the arguments above are relevant to subAlfvenic turbulence. They are suggestive that the gradients of the velocities and magnetic fields are aligned with the {\it local} magnetic field and therefore sample the local direction of the magnetic field flux at the largest of the two scales, one is being the turbulence dissipation scale, the other is the telescope resolution scale. 
This point is very important for the technique that we are going to propose. In this paper we are claiming that by measuring the magnetic field gradients one can trace the turbulent magnetic-field in the volume under study.

The discussion above was focused on the Alfvenic turbulence. In incompressible conducting fluid in 3D, apart of Alfven modes, the pseudo-Alfven modes exist (see GS95). The latter is the limiting case of slow modes in the incompressible limit. Pseudo-Alfven modes and, in general, the slow modes are slaved by Alfvenic modes, that shear them both in the case of magnetically dominated (low-$\beta$) and gas-pressure dominated (high-$\beta$)
plasmas (GS95, \citealt{Lithwick2001CompressiblePlasmas, Cho2002CompressiblePlasmasb, Cho2003CompressibleImplicationsb, Kowal2010VelocityScalingsb}). Thus we expect that the slow modes will also show the properties of the magnetic gradients similar to the Alfven waves.

The third fundamental MHD turbulent mode is the fast mode. The fast modes are different from both Alfven and slow modes. The fast modes create an accustic-type cascade \citep{Cho2002CompressiblePlasmasb, Cho2003CompressibleImplicationsb} which marginally cares about magnetic-field direction. In terms of our attempts to use gradients to trace magnetic fields, fast modes are distort the alignment. Therefore, it is important that numerical simulations indicate that the fast modes
are subdominant even for supersonic driving \citep{Cho2002CompressiblePlasmasb}. Therefore having a natural admixture of Alfven and slow modes, which is augmented by fast modes and shocks, we expect to see alignment of magnetic gradient perpendicular to the local magnetic-field direction. This is the theoretical conclusion that motivates our study below. 

We may add that for the weakly compressible flows the density associated with slow waves will mimic the GS95 scalings \citep{Beresnyak2005DensityTurbulence}. However, for supersonic flows the production of shocks significantly disturbs the statistics of density. As a result, for subsonic flows density gradients are also expected to be aligned perpendicular to the magnetic field which explains the results in empirical results in \citep{Soler2013} as well
as our numerical experiments with density gradients in \citeauthor{GL17} (\citeyear{GL17}, henceafter GL17) and \citeauthor{YL17} (\citeyear{YL17}, henceforth YL17). In terms of magnetic field tracing the density gradients are expected to be inferior to velocity and magnetic field gradients, 
which is clearly shown in \cite{YL17b} where the GALFA HI intensity gradients are compared to both and velocity centroid gradients and magnetic fields as revealed by Planck data. At the same time, the density gradients can also be very important. For instance, the misalignment of density gradients and magnetic field directions can be informative in terms of shocked gas and supersonic flows.

A note about observational availability of magnetic field/velocity gradients is due. The observations probe the properties of diffuse media along the line of sight, rather than at a point in 3D space. Therefore one can wonder about the contributions from different scales that affect {\it observationally available} gradients. The magnetic gradients are can be estimated as the ratio $|b(r_1)-b(r_2)|/|r_1-r_2|$, where $b(r)$ is the magnetic field at the point $r$. The structure function $d(r)=\langle (b(r_1)-b(r_2))^2\rangle$ can give us some indirect insight into the process of the summation along the line of sight. As we discuss in the next section (see Eq. (\ref{synch})) , the observed intensities are proportional to $\int_0^D dz H_{\bot}^2 $, where the integration is performed along the line of sight through the diffuse volume of thickness $D$. Structure functions of synchrotron intensities for the general case of anisotropic turbulence were studied in detail in \cite{LP12}. This function, i.e. $S(l)=\langle (I(e_1)-I(e_2))^2 \rangle$, where $l$ is the distance between the lines of sight, is roughly proportional to
\begin{equation} 
\int [d((l^2+z^2)^{1/2})-d(z)] dz, 
\label{structure}
\end{equation}
where the integration is performed along the line of sight. For the Kolmogorov-type turbulence it is possible to show that the major contribution to the integral is coming from the scales close to $l$. 

Magnetic and velocity perturbations are symmetric within Alfvenic turbulence. However, the differences are obvious in compressible flows. For instance, shocks distort the velocity structure creating velocity gradients parallel to the local direction of magnetic field. This effect is not present for magnetic field, which gradients are more robust to compressible turbulence. Similarly, the velocity gradients for the flows affected by gravitational collapse tend to be parallel to the ambient magnetic field \citep{YL17} This is not expected for magnetic field gradients making them a more robust measure of the line of sight integrated magnetic field.  

\subsection{Synchrotron gradient properties}

Synchrotron emission arises from relativistic electrons spiraling about magnetic fields (see \citealt{1970ranp.book.....P}  references therein). A quantitative study of the synchrotron emission (see \citealt{1959ApJ...130..241W}) revealed that the emission is non-linear in the magnetic field $H$ with the origin of nonlinearity arising from relativistic effects. For the power law distribution of electrons $N(E) dE\sim E^{\alpha} dE$, the synchrotron emissivity is
\begin{equation}
I_{synch}(x, y) \propto \int dz H_{\bot}^{\gamma} (x, y, z),
\label{synch}
\end{equation}
where $H_{\bot}=\sqrt{H_x^2+H_y^2}$ corresponds to the magnetic field component perpendicular to the line of sight, the latter given by the z-axis. The fractional power of the index $\gamma= (\alpha +1)/2$ 
was a impediment for quantatitive synchrotron statistical studies. However, the problem of the magnetic field dependence on the fractional power was dealt with in \cite{LP12}, where it was shown that the correlation functions and spectra of $H_{\bot}^{\gamma}$ can be expressed as a product of a known function of $\gamma$ times the statistics of $H_{\bot}^2$, i.e. the synchrotron intensity obtained for $\alpha=3$. Although we do not use the correlation function approach explicitly, our approach is based on the statistical properties turbulence and we expect that similar to the case considered in \cite{LP12} the gradients calculated with $\alpha=3$ will correctly represent the results for other $\gamma$. Thus the fractional power of magnetic field in Eq. (\ref{synch}) will not be considered as an issue within the present study aimed at determining magnetic field gradients. 

In Eq. (\ref{synch}) we disregard the dependence on the relativistic cosmic electron density. This is justified as we are interested by the gradients at the small scales at which the distribution of relativistic electrons is smooth in most parts of the diffuse media. In other words, on the basis of what we know about cosmic ray propagation (see e.g. Lazarian \& Yan 2014 and ref. therein) we expect that the gradients arising from the inhomogeneities of the relativistic electrons distribution to be subdominant to the gradients arising from turbulent magnetic fields. 

Alfven modes do not change the strength of magnetic field. Therefore if the line of sight is directed along the $z$-axis, while the
the Alfven mode is polarized in the $x-y$ plane, the magnetic field fluctuations are perpendicular to the line of sight and therefore the observed synchrotron intensity does not change with the amplitude of the Alfven mode. However, if all 3 components of the Alfven mode are present, then the fluctuations in the z-direction result in the decrease of the observed synchrotron intensities (see LP12). Therefore while the magnetic field gradients are still in the direction perpendicular to the local magnetic field, the synchrotron intensity fluctuations are parallel to the local magnetic field direction. The statistical similarity of the Alfven and slow modes (see \citealt{LP12}) is suggestive that the slow modes behave the same way. At the same time, fast modes are expected to play a disruptive role for the magnetic field tracing. For instance, fast modes in magnetically dominated plasma correspond to the compressions of magnetic field. These compressions happen perpendicular to magnetic field and therefore the fluctuations of synchrotron intensity induced by fast modes are expected to be perpendicular the magnetic field directions. For the pressure dominated media, the so-called high $\beta$ media, where $\beta$ is the ratio of the gas to magnetic pressure, the fast modes are essentially sound waves and they only marginally compress magnetic field (see GS95, Cho \& Lazarian 2003). This mitigates the effect of fast modes for the gradient technique in that we introduce in this paper.  

\section{Numerical data}

We test our theory using numerical simulations obtained from three different codes, in particular, from a 3D MHD compressible code described in \citealt{Cho2002CompressiblePlasmasb}), an imcompressible code described in  \cite{Cho2000TheTurbulence}, as well as the data sets used in YL17 and its subsequent studies from another 3D MHD gravity-supported compressible code family ZEUS-MP. The use of numerical results obtained with different codes is advantageous and this allows us to test better the gradient technique for various physical settings. For instance, the 3D MHD compressible code from \cite{Cho2002CompressiblePlasmasb} is a third-order accurate hybrid, that employs essentially non-oscillatory (ENO) scheme on solving the ideal isothermal MHD equations in a periodic box. For our case, we choose $M_s=0.5,3.0,10.0$ and $M_A=0.1$ for our purpose. The code from \cite{Cho2000TheTurbulence}, on the other hand, solves the periodic incompressible MHD equations using the pseudo-special code. The resultant data corresponds to the extreme case plasma $\beta = 2M_A^2/M_s^2 = \infty$. In our case, we used an incompressible cube with $M_A=0.80$. The respective parameters are listed in Table \ref{tab:simulationparameters}. We then follow \cite{2016ApJ...831...77L} to produce both the maps of synchrotron polarization and synchrotron intensity.
 
\begin{table*}[h]
 \centering
 \label{tab:simulationparameters}
 \caption {Simulations used in our current work.}
 \begin{tabular}{c c c c c c}
  Code &  Model & $M_s$ & $M_A$ & $\beta=2(\frac{M_A}{M_s})^2$ & Resolution\\ \hline \hline
  ZEUS-MP family &  b11 & 0.4 & 0.04 & 0.02 & $480^3$\\\hline
  \begin{comment}
 &   b12 & 0.8 & 0.08 & 0.02 & $480^3$\\
 &  b13 & 1.6 & 0.16 & 0.02 & $480^3$\\
 &  b14 & 3.2 & 0.32 & 0.02 & $480^3$\\
 &  b15 & 6.4 & 0.64 & 0.02 & $480^3$\\
 &  b21 & 0.4 & 0.132 & 0.2178 &$480^3$\\
 &  b22 & 0.8 & 0.264 & 0.2178 &$480^3$\\
 &  b23 & 1.6 & 0.528 & 0.2178 &$480^3$\\
 &  b31 & 0.4 & 0.4 & 2 & $480^3$\\
 &  b32 & 0.8 & 0.8 & 2 & $480^3$\\
 & b41 & 0.132 & 0.4 & 18.3654 & $480^3$\\
 &  b42 & 0.264 & 0.8 & 18.3654 & $480^3$\\
 & b51 & 0.04 & 0.4 & 200 & $480^3$\\
 &  b52 & 0.08 & 0.8 & 200 & $480^3$\\\hline
\end{comment}
 \cite{Cho2002CompressiblePlasmasb} & b.5 & 0.5 & 0.1 & 0.08 & $512^3$ \\
 & b3 & 3.0 & 0.1 & 0.0022 & $512^3$ \\
 & b10 & 10.0 & 0.1 & 0.0002 & $512^3$ \\ \hline
 \cite{Cho2000TheTurbulence} & Incompressible & 0 & 0.8 & $\infty$ & $512^3$ \\ 
  & Incompressible & 0 & 3.2 & $\infty$ & $512^3$ \\ \hline\hline
\end{tabular}
\end{table*}

%In Figure \ref{fig:0} we illustrate We test our theory numerical simulations obtained with our 3D MHD compressible code (see more details in \citealt{Cho2003CompressibleImplicationsb}) and following the procedures described in \cite{2016ApJ...831...77L} created  To test what happens in the limit of incompressible turbulence we also use the results of simulations obtained with 3D incompressible MHD code (see \citealt{Cho2000TheTurbulence}). 
  
\section{Properties of Synchrotron Intensity Gradients (SIGs)}
\label{sec:3}

\subsection{Calculation of SIGs}

To calculate Synchrotron Intensity Gradients (SIGs), we use the procedure for gradient calculation that we introduced in YL17. The procedure consists of three steps. We first pre-process our synchrotron intensity maps with an appropriate noise-removal Gaussian filter. We then interpolate the map to ten times its original level, and determine the gradient field by computing the maximum gradient direction in the interpolated synchrotron intensity maps. By probing the peak in gradient orientation distributions in the sub-blocks of the gradient map, we gain an estimate of the {\it sub-block averaged} gradient vector as in YL17. That allows us to compare our magnetic field predictions to those revealed by the generally accepted technique of tracing polarization. As discussed in YL17, the sub-block averaging approach provided a way of estimating how good the vector is being predicted  in a block: The gradient angle distribution peaks tells the predicted value, while the shape of the distribution tells how good the prediction is. The deviation from the Gaussian function provides an error estimate for us to judge whether our method is accurate within a block.

\subsection{SIGs from Alfven, Slow and Fast modes}

To illustrate the applicability of our theoretical considerations related to synchrotron intensity gradients by providing the comparison of the SIG, we perform the mode decomposition that similar to that in \cite{Cho2002CompressiblePlasmasb,Cho2003CompressibleImplicationsb}.  The corresponding equations determining the basis for the decomposition into modes are:
\begin{subequations} \label{eq:fsa-decompositions} 
\begin{align} 
\hat{\zeta}_f &\propto (1+\beta+\sqrt{D}) (k_\perp \hat{\bf{k}}_\perp) +(-1+\beta+\sqrt{D}) k_{||}\hat{\bf{k}}_{||} \\
\hat{\zeta}_s &\propto (1+\beta-\sqrt{D}) (k_\perp \hat{\bf{k}}_\perp) +
(-1+\beta-\sqrt{D}) k_{||}\hat{\bf{k}}_{||}\\
\hat{\zeta}_f &\propto -\hat{\bf{k}}_\perp\times \hat{\bf{k}}_{||}
\end{align}\end{subequations}
where $D=(1+\beta/2)^2-2\beta \cos\theta$, $\beta=\frac{\bar{P}_g}{\bar{P}_B}=\frac{2M_A}{M_s}$, and $\cos\theta= \hat{k}_{||} \cdot \hat{B}$. 
We would only use the LOS component of the decomposed velocities for magnetic field calculations.That is to say, the three velocity modes can then be acquired by
\begin{equation}
b_{(f,s,a),z}= [\mathscr{F}^{-1}(\mathscr{F}(\bf{b})\cdot\hat{\zeta}_{f,s,a})](\hat{\zeta}_{f,s,a} \cdot \hat{\zeta}_{LOS})
\end{equation}
where $\mathscr{F}$ is the Fourier transform operator. 

\begin{figure*}[tbhp]
\centering
\includegraphics[width=0.99\textwidth]{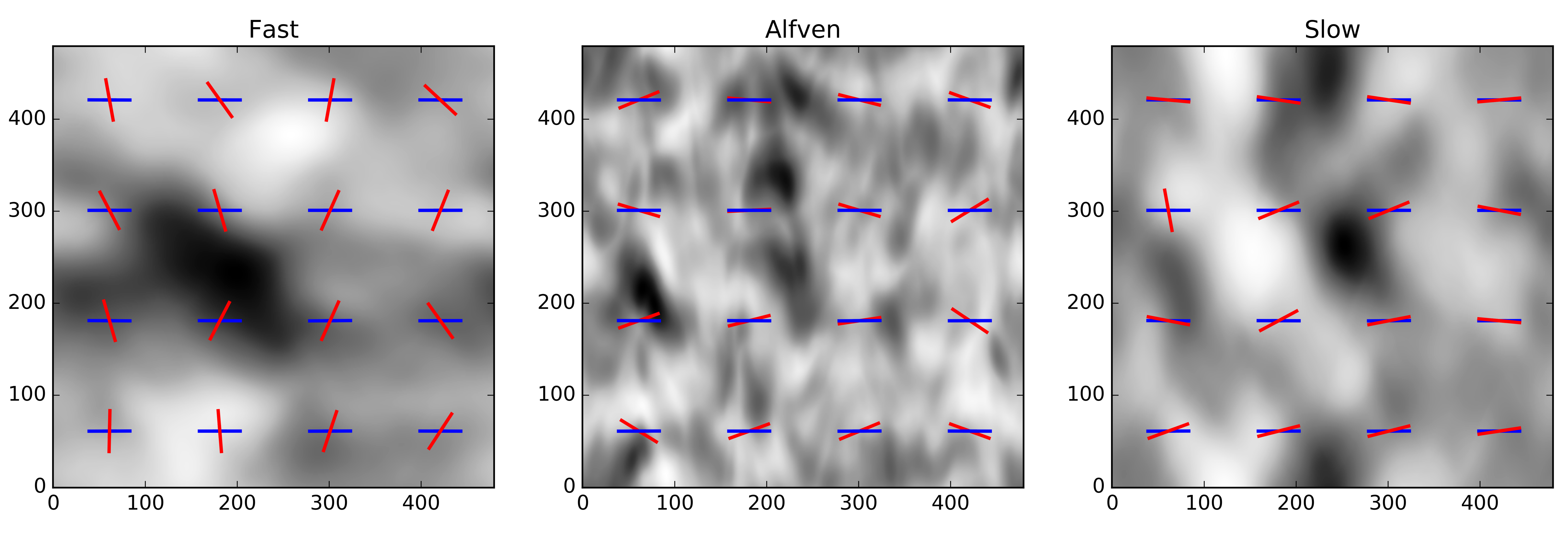}
\caption{\label{fig:0} Illustration on the synchrotron intensity maps after wave-mode decomposition on magnetic field, following \cite{Cho2003CompressibleImplicationsb} on our cube b11. Red vectors are SIGs, while blue vectors are synchrotron polarization. It is obvious that only Alfven and slow modes contributes to the alignment of SIGs to magnetic field. }
\end{figure*}
Figure \ref{fig:0} exhibits synthetic observations produced with the separated MHD modes from one of the cube in the ZEUS family simulation. It shows that, as we expected, that the SIGs are aligned parallel to magnetic field for the case of Alfven and slow modes. At the same time, also as expected, for the case of magnetically dominated media, the SIGs are perpendicular to the plane-of-sky projected component of the magnetic field. In our simulations the ratio of the Alfven, slow and fast modes is 1:0.7:0.3. Therefore, indeed, the effect of the fast modes is subdominant and we expect that for the actual simulations without any decomposition the SIGs will trace magnetic field. This is what we test below.

\subsection{SIGs: Effect of Block size}

\begin{figure*}[h]
\centering
\includegraphics[width=0.33\textwidth]{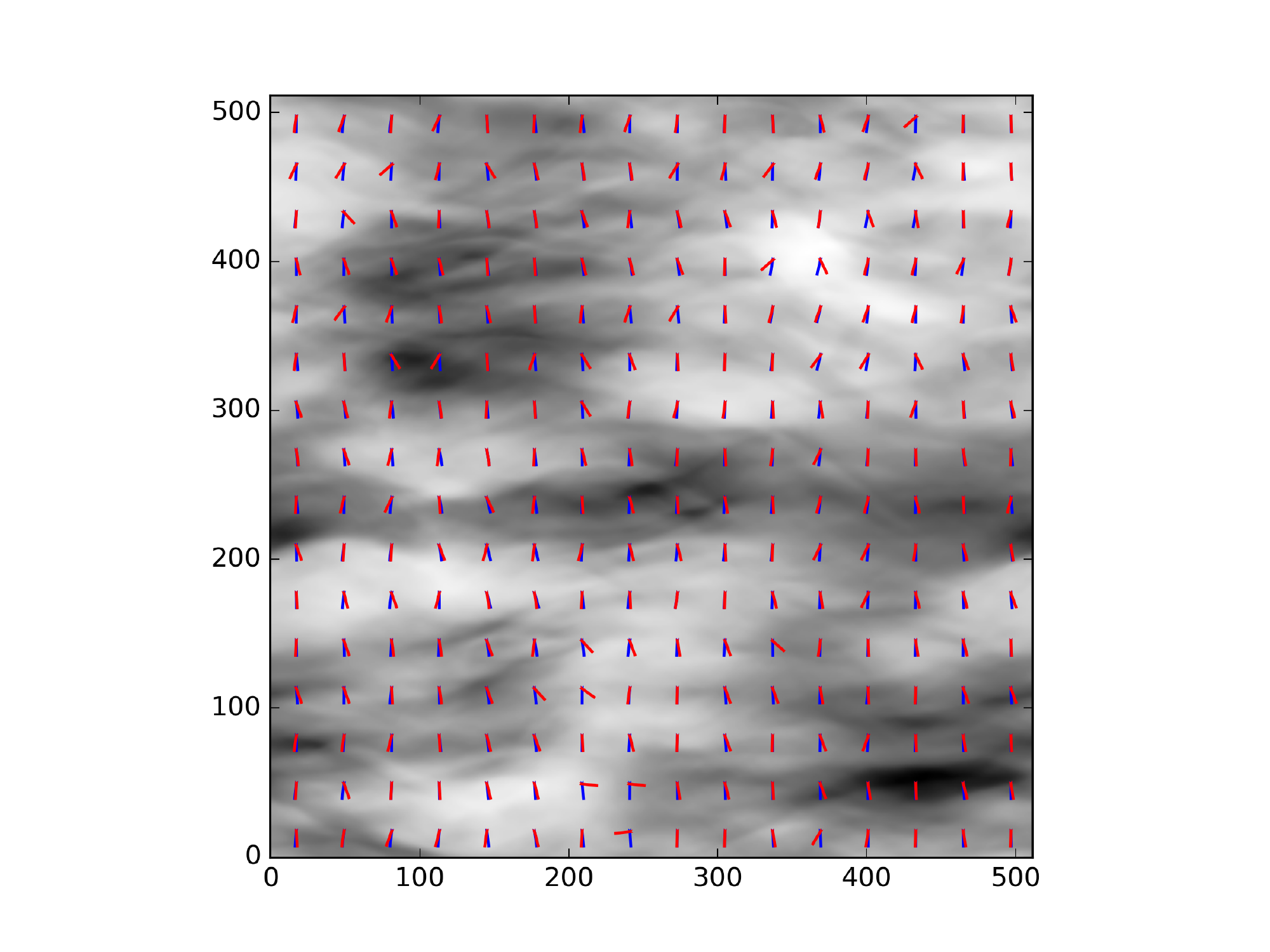}
\includegraphics[width=0.33\textwidth]{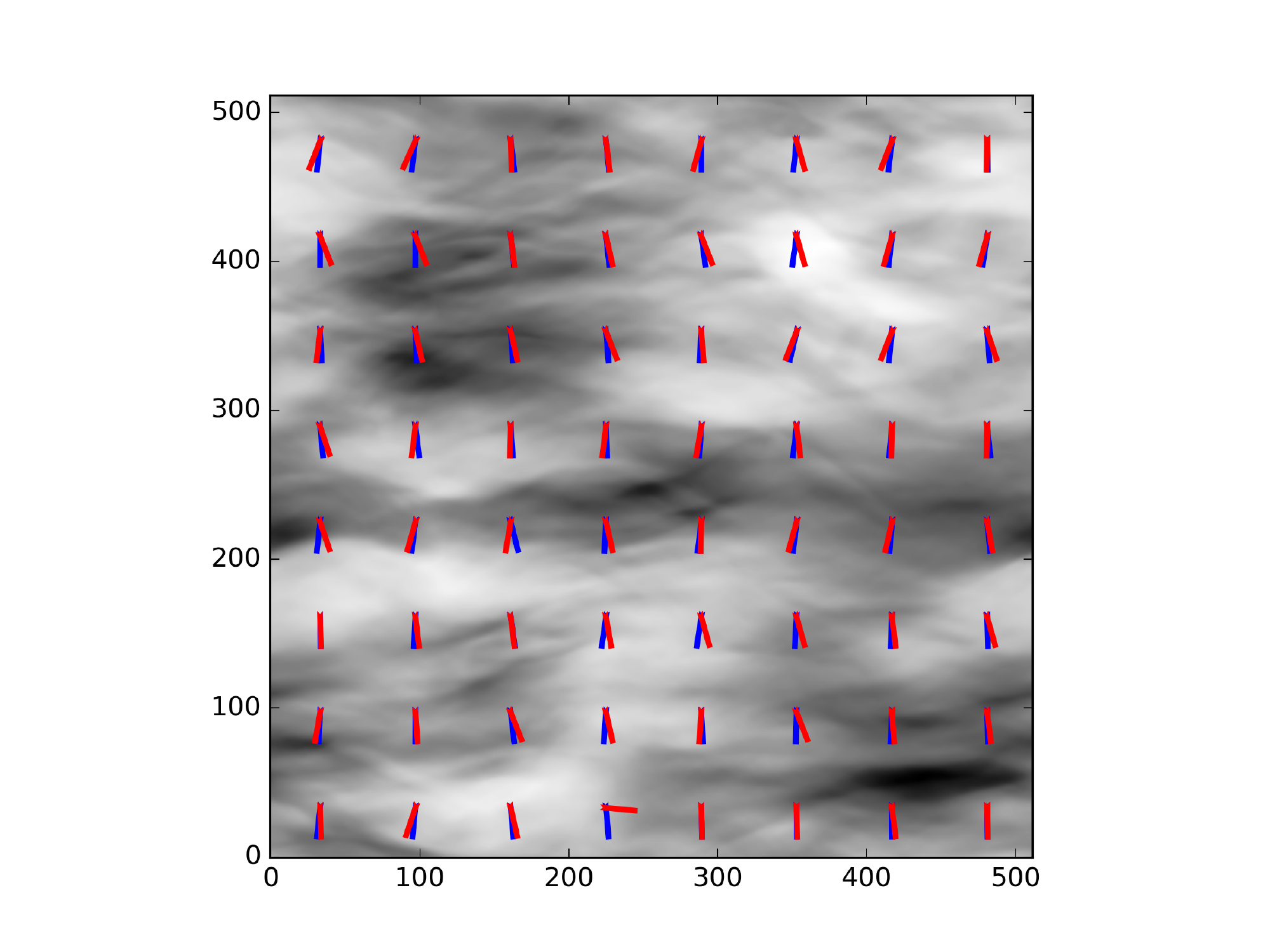}
\includegraphics[width=0.33\textwidth]{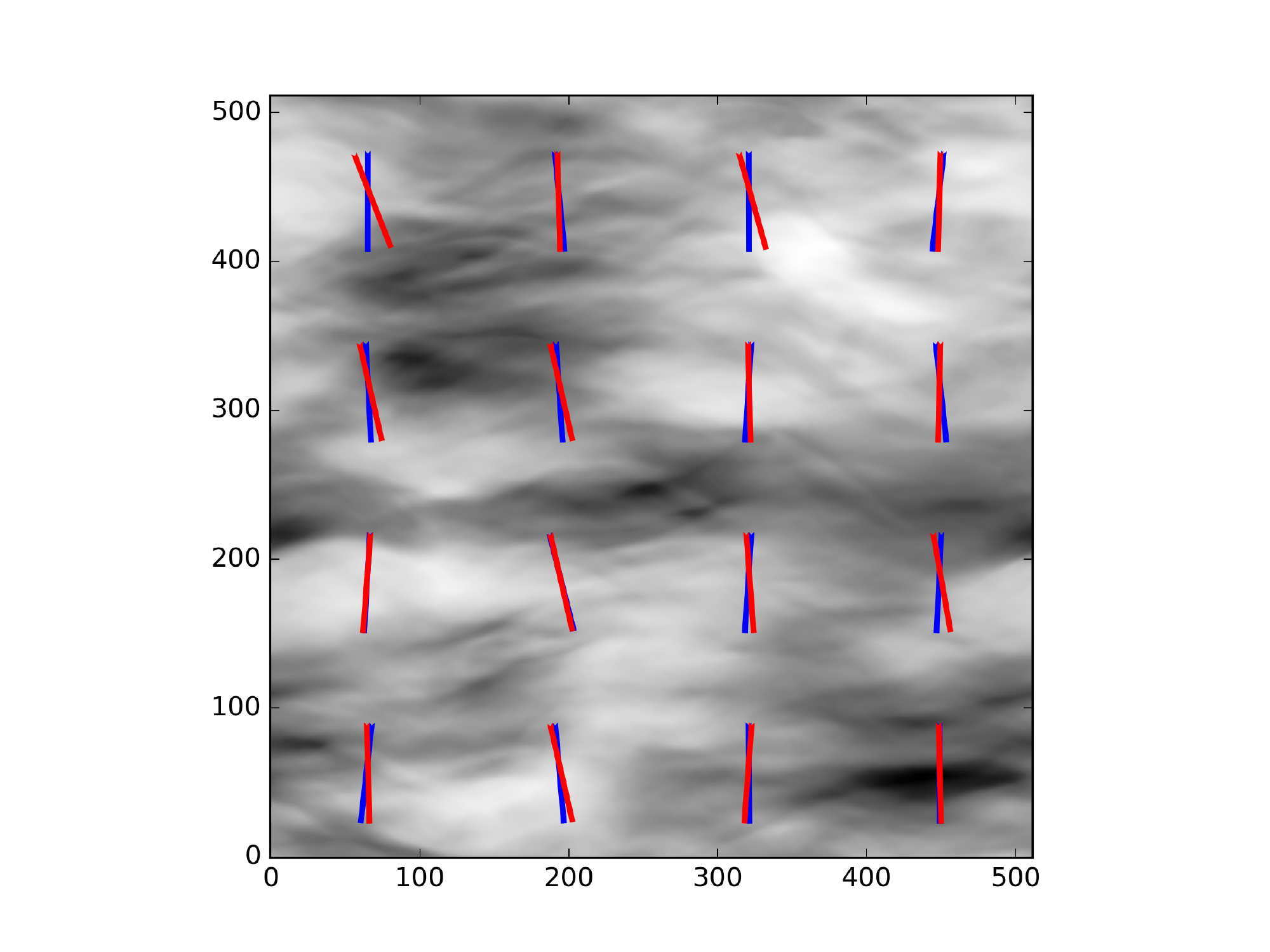}
\includegraphics[width=0.33\textwidth]{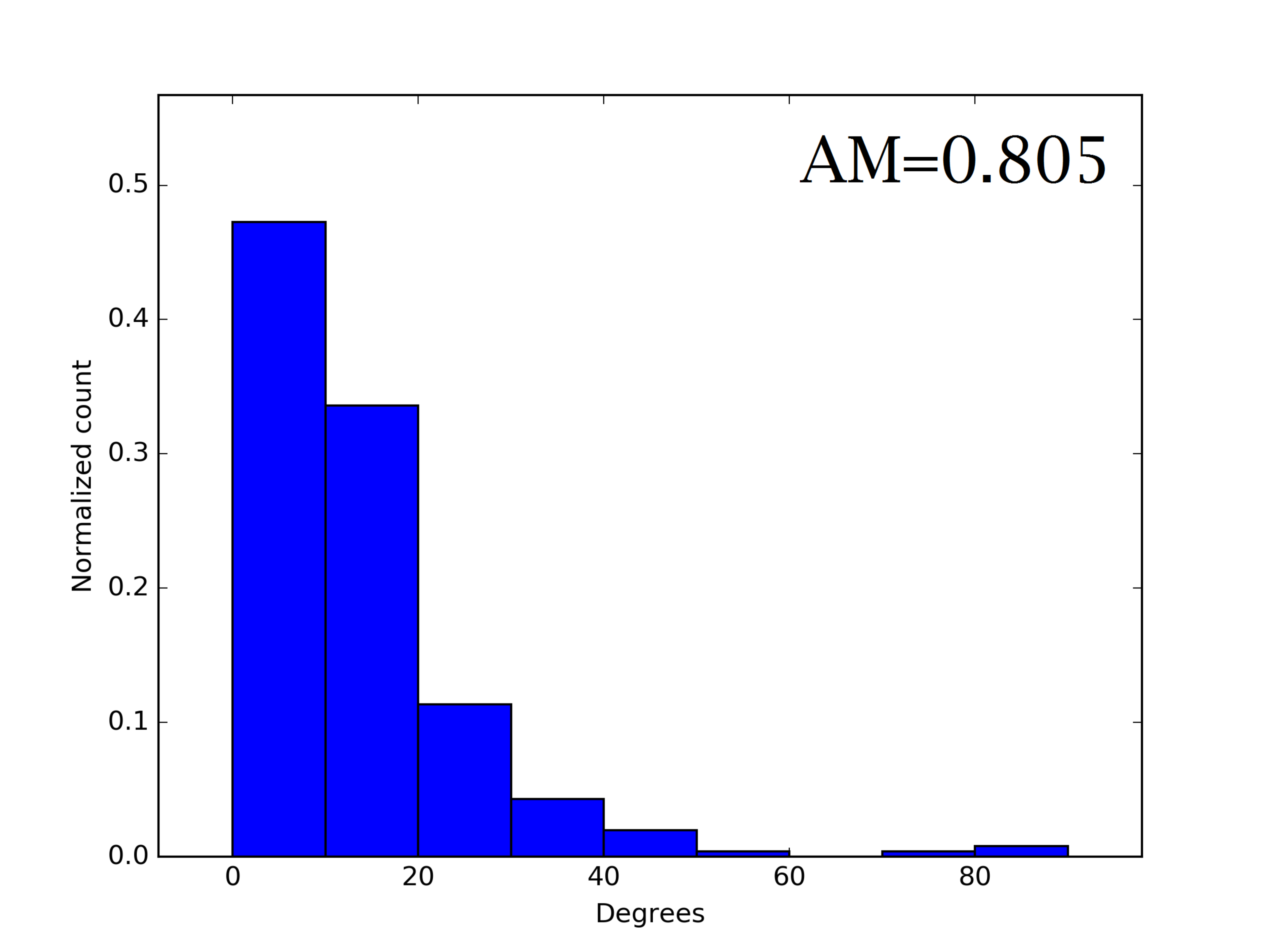}
\includegraphics[width=0.33\textwidth]{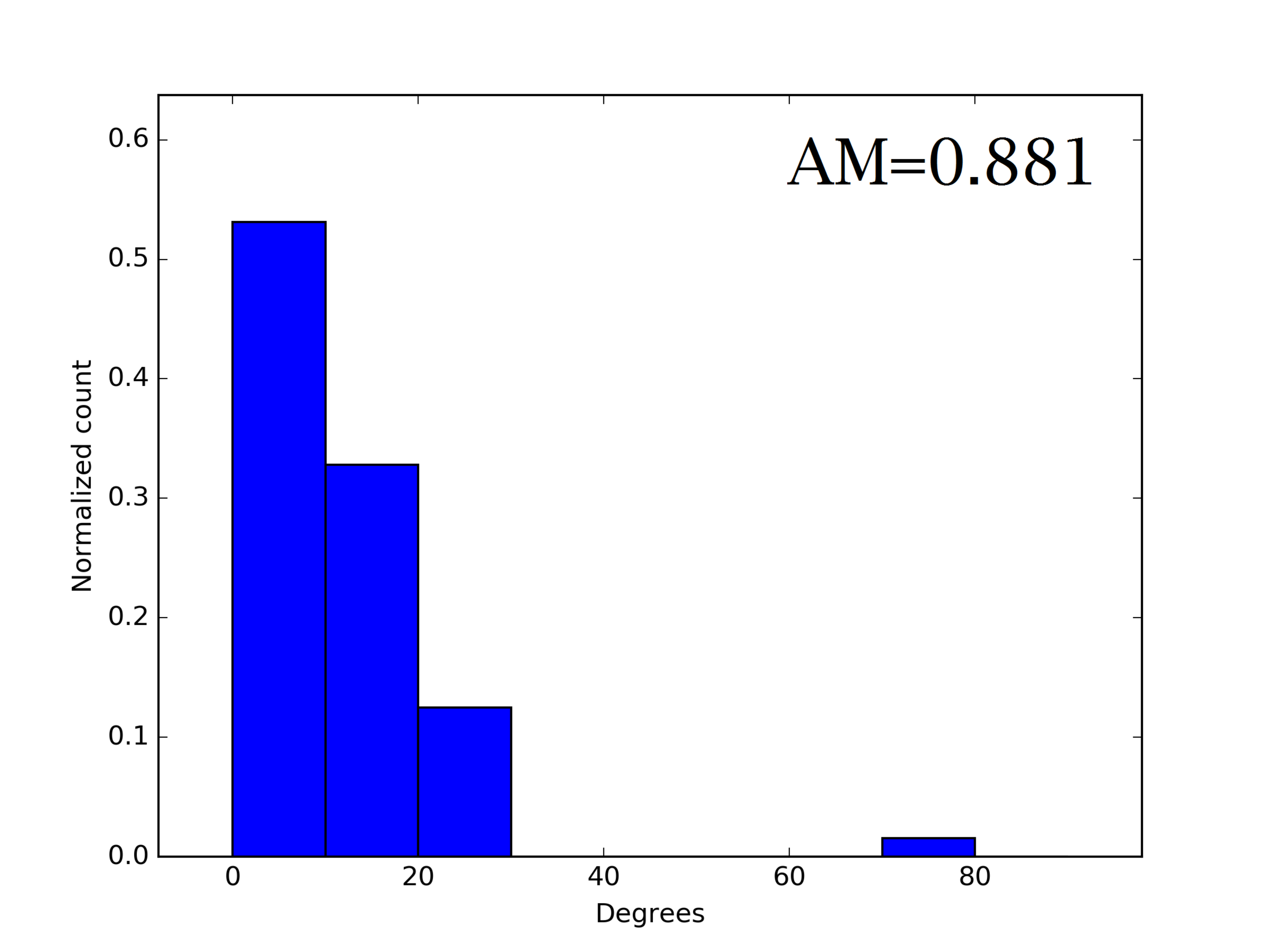}
\includegraphics[width=0.33\textwidth]{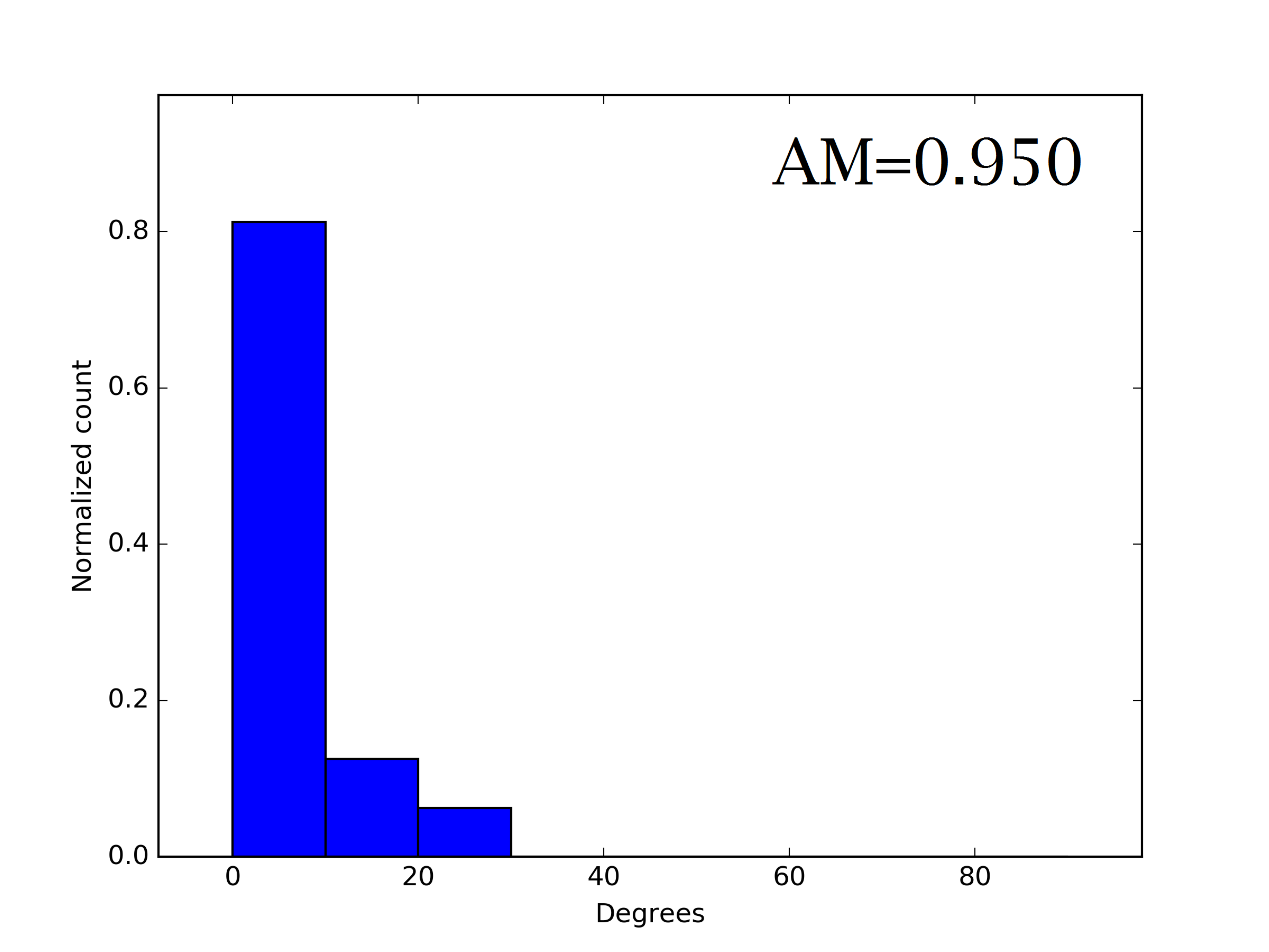}
\caption{\label{fig:1} SIGs (red) and synchrotron polarization (blue) from one of our $512^3$ MHD simulation with $M_s=0.5$ and $M_A=0.1$, but with different block sizes: 32 (left), 64 (middle), 128 (right), which corresponds to the physical conditions that emit most of the synchrotron radiation, covering on the synchrontron intensities on each setting, respectively. Larger averaging corresponding to bigger blocks provides a more accurate measurements of magnetic-field direction. The histograms below each plot are the relative angle distributions between the SIGs to magnetic field}
\end{figure*}

Figure \ref{fig:1} demonstrates that out approach can deliver SIGs in a robust way with the magnetic-field directions obtained with SIGs providing an adequate representation of the projected magnetic field. To demonstrate the latter point in Figure \ref{fig:1} we also show the magnetic field directions as traced by the synchrotron polarization in the synthetic observations. 

To quantify how well the SIGs are tracing the synchrotron polarization that represents the projected magnetic field we introduce the alignment measure:
\begin{equation}
 AM=2\langle \cos^2\theta-1 \rangle,
 \label{AM}
 \end{equation} 
where $\theta$ is the angle between the SIG direction and magnetic field direction derived from Planck synchrotron polarization measurements, which $AM \in [-1,1]$.  When $AM=1 $, that indicates a perfect alignment between SIGs to magnetic field. When it becomes zero, there are no relation between SIGs and magnetic field. When $AM=-1$, SIGs tend to be perpendicular to magnetic field. 

\begin{figure*}[tbhp]
\centering
\includegraphics[width=0.96\textwidth]{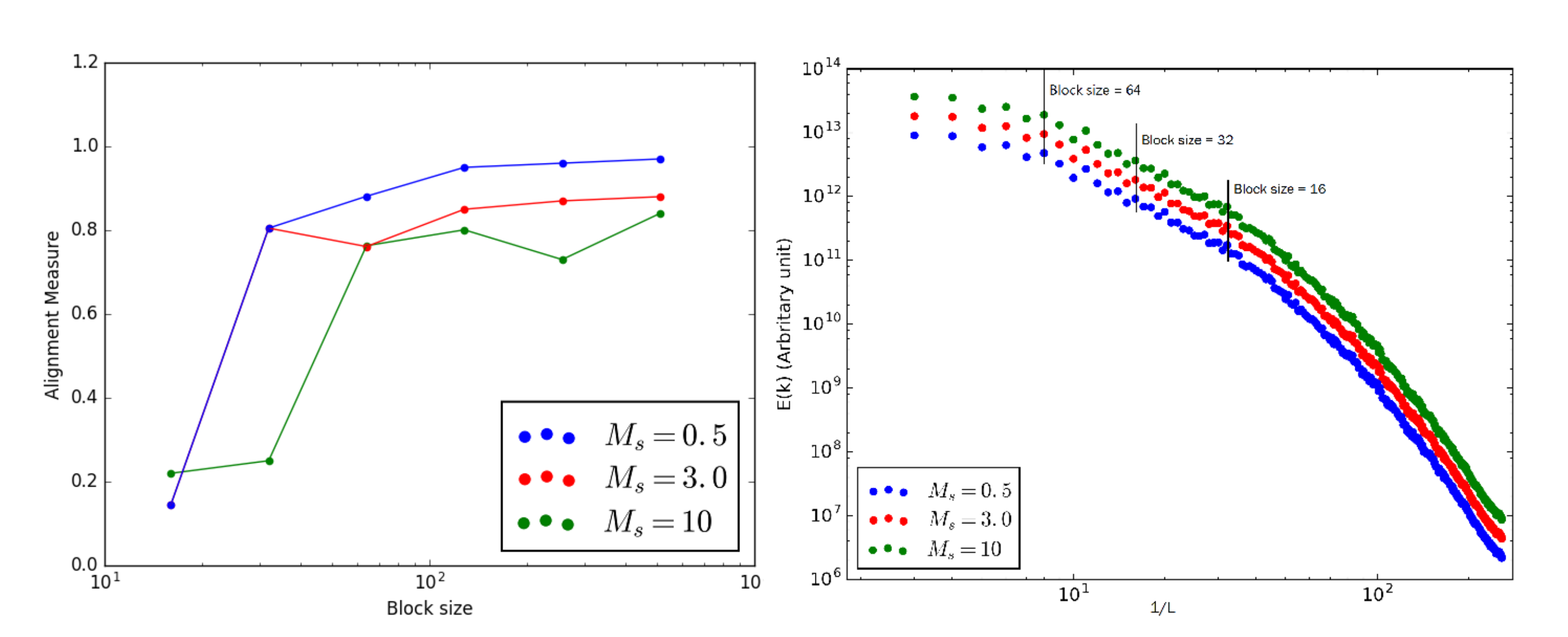}
\caption{\label{fig:block-AM} (Left) The variation of AM to block size for our three different synthetic synchrotron maps. (Right) The respective spectrums of the three systems.}
\end{figure*}

The alignment between the SIGs and the magnetic field increases with the block size, but we observe a relatively small increase starting with a particular block size. This effect is demonstrated by the the left-hand-side  panel of Figure \ref{fig:block-AM} that shows that starting with the block size $\sim 64$ pixels the increase of the $AM$ with the block size get very slow . Observationally this size provides the optimal block size for the analysis that maximizes the informational output of the SIG technique. Note, that the rapid decrease of the alignment measure as the block size decreases is strongly affected by the numerical resolution. Our analysis of the spectral slope of the turbulence on the right of Figure \ref{fig:block-AM} indicates that starting with scales $k \sim 40$, which is about $r = 512/40 = 12.8$ pixels,  the structures that we see are dominated by the numerical effects. For the part of the turbulent cascade that is dominated by numerical effects we do not expect to observe the GS95 or weak turbulence scalings of the turbulent magnetic field. Therefore it is not surprising that the SIG technique fails. In fact, this is in a good agreement with the fact that for block size less and $\sim 16$ we do not see good alignment, as shown in the left of Figure \ref{fig:block-AM}. At the same time, this is suggestive that for the actual low-noise astronomical observations the size of the optimal block may be smaller than 64 pixels that we find in our numerical simulations. Indeed, unlike numerical simulations, the astrophysical turbulence exhibits an extensive inertial range with the dissipation scale too small to be resolved by observations (see Chepurnov \& Lazarian 2009). 
 
\subsection{SiGs: effect of the sonic and Alfven Mach numbersr}

\begin{figure*}[tbhp]
\centering
\includegraphics[width=0.33\textwidth]{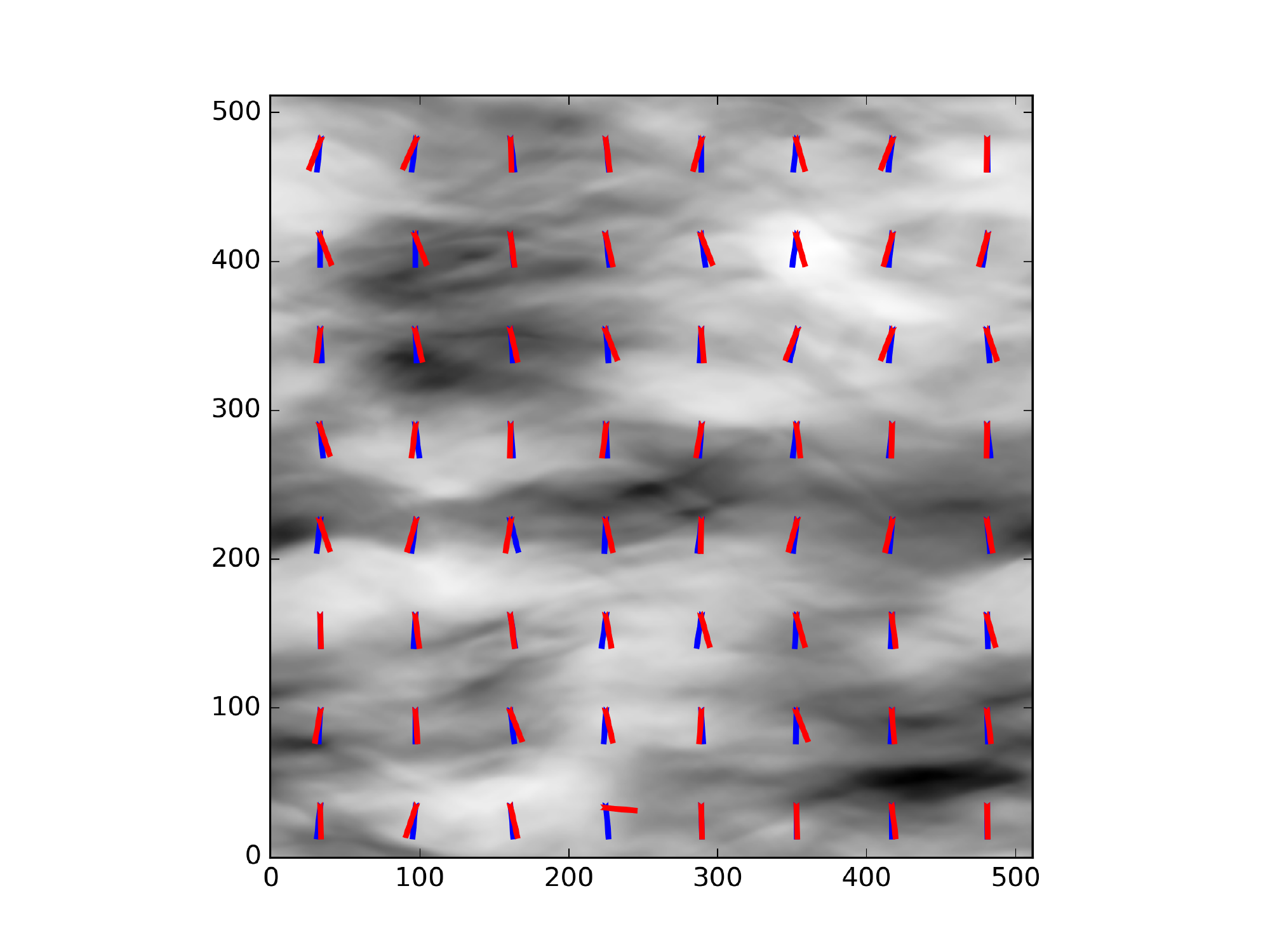}
\includegraphics[width=0.33\textwidth]{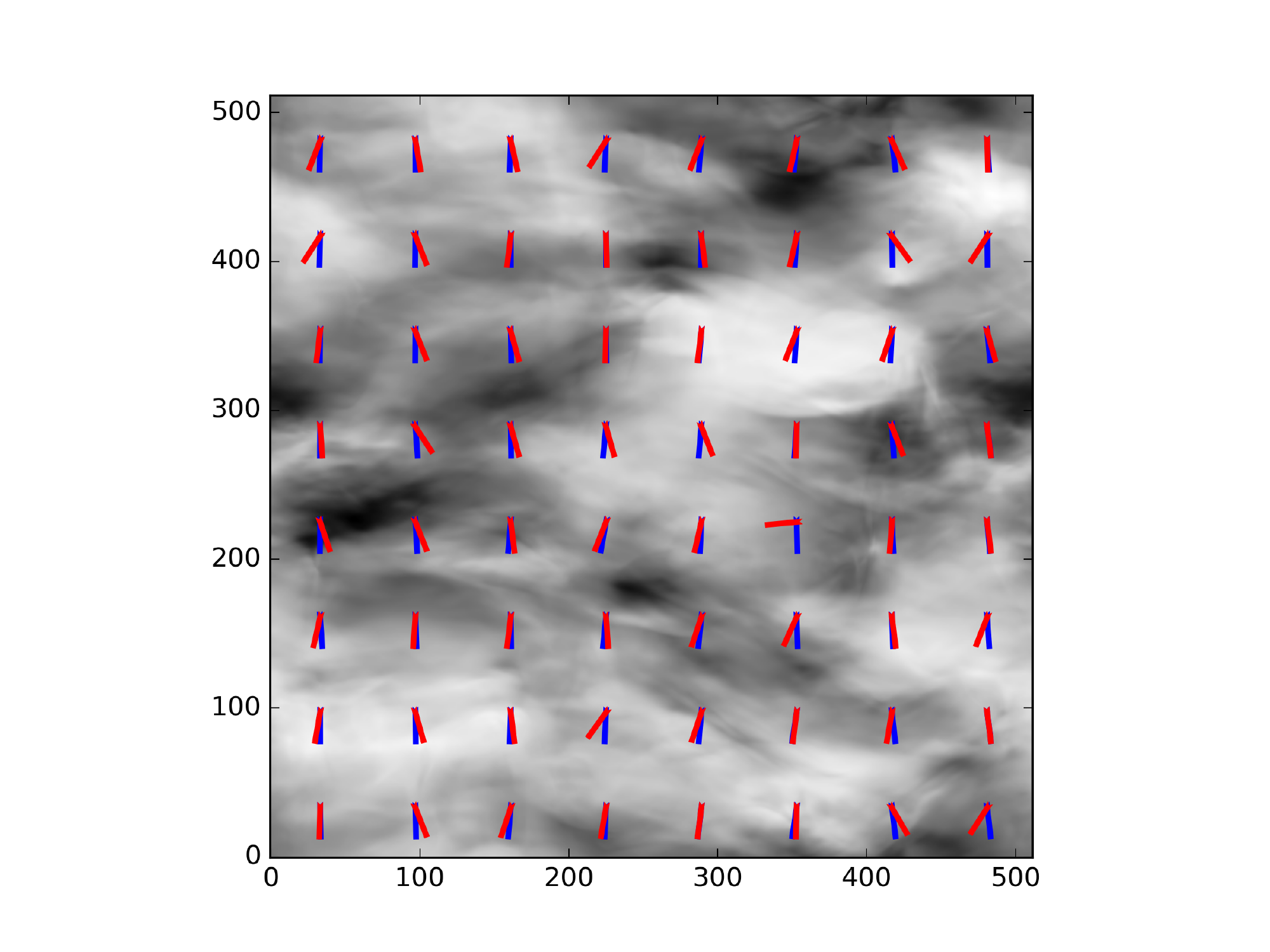}
\includegraphics[width=0.33\textwidth]{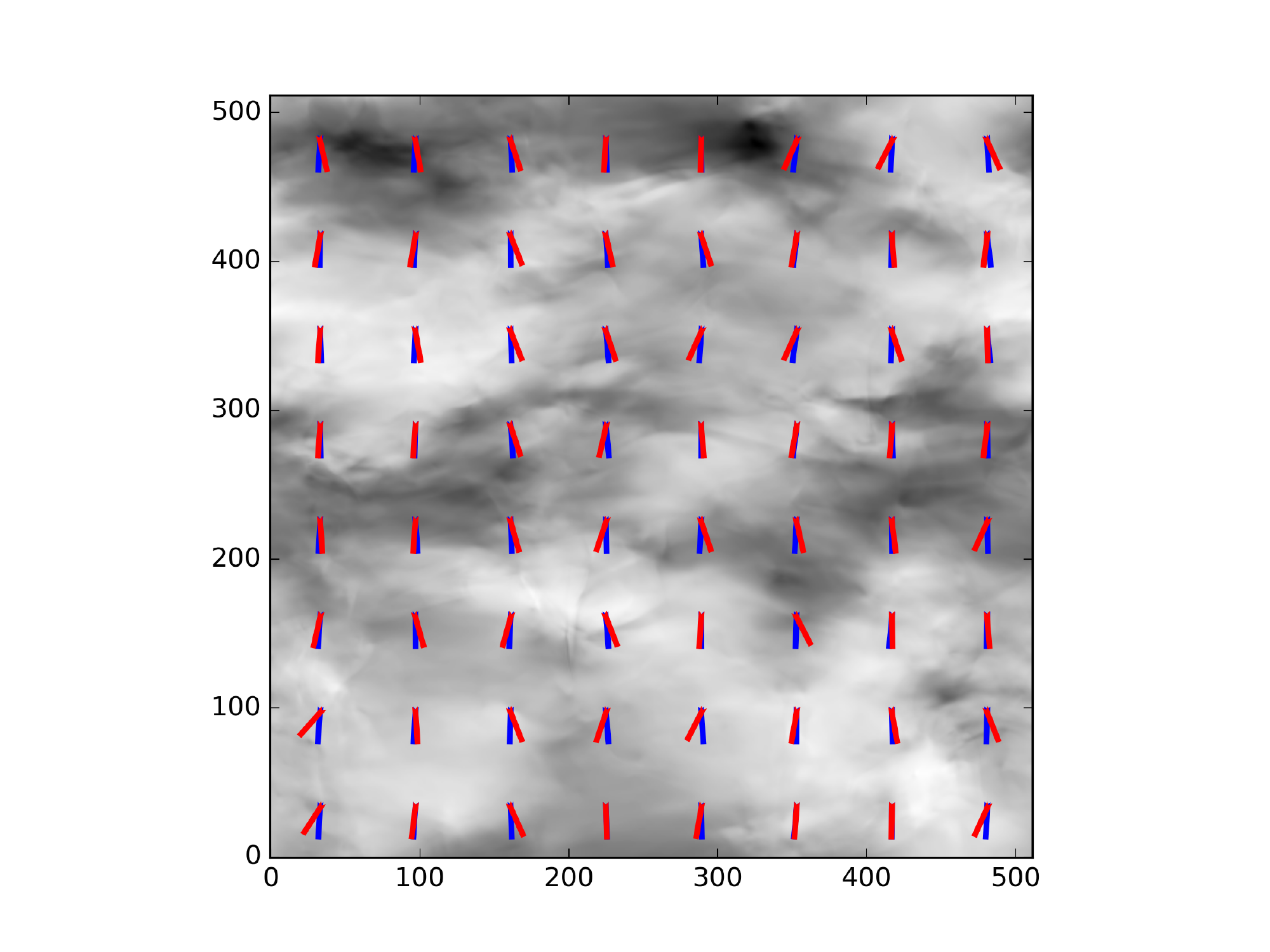}
\includegraphics[width=0.33\textwidth]{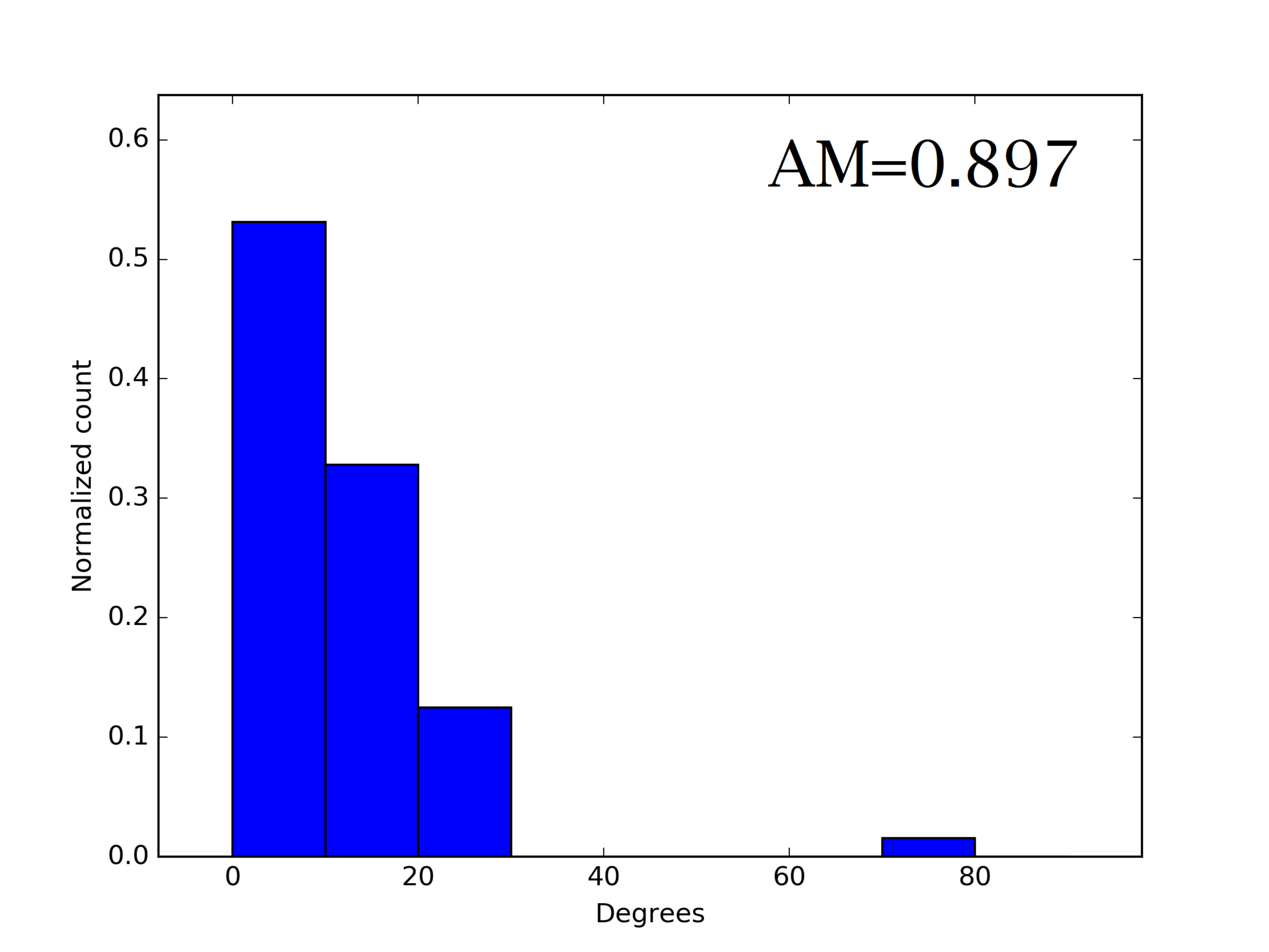}
\includegraphics[width=0.33\textwidth]{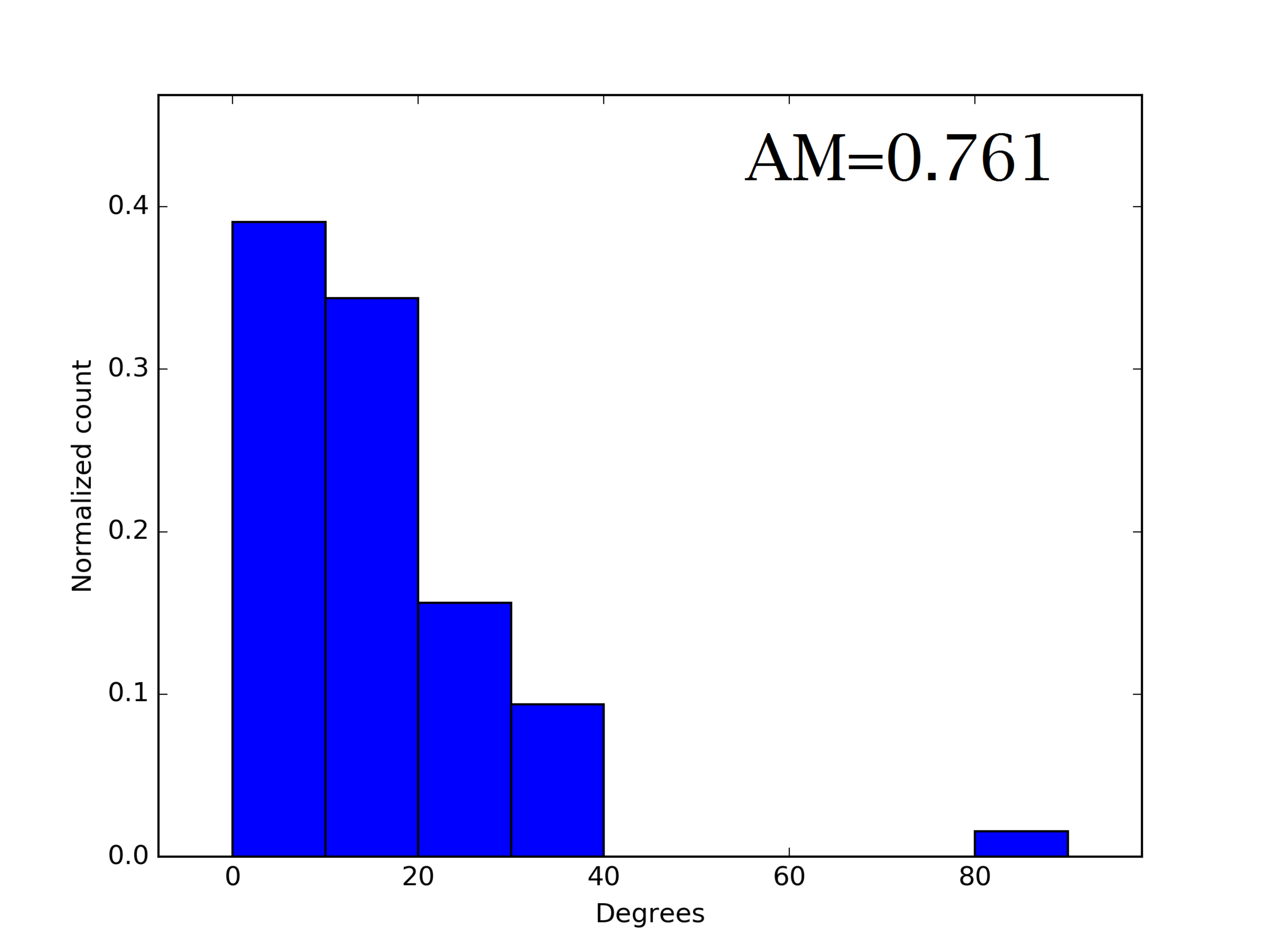}
\includegraphics[width=0.33\textwidth]{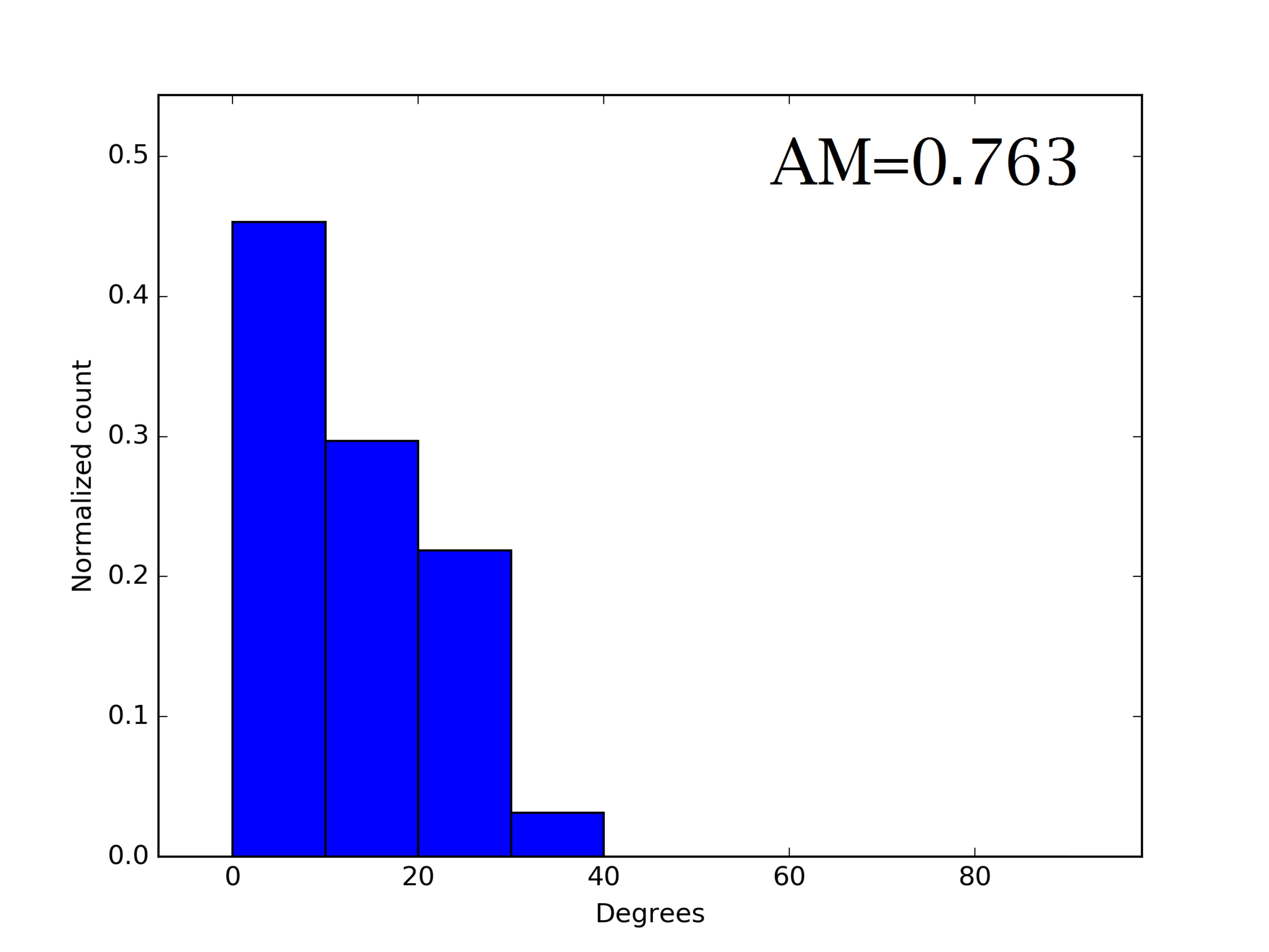}
\caption{\label{fig:2} SIGs (red) and synchrotron polarization (blue) from three sets of turbulent trans-Alfvenic MHD simulations with sonic Mach numbers $M_s=0.5$ (left), $3.0$ (middle) and  $10.0$ (right), respectively. The background shows the synchrotron intensities. In the histogram below each panel shows the relative angle between SIG and magnetic field directions.  The histograms below each plot are the relative angle distributions between the SIGs to magnetic field}
\end{figure*}

For most of the environments of the spiral galaxies the areas dominating the synchrotron emission may correspond to the hot gas with low sonic Mach numbers $M_s$. However,  it is interesting to explore to what extend the effects of compressibility can affect the SIG technique. Therefore test how the SIGs trace magnetic field in systems with different $M_s$.The upper panels of Figure \ref{fig:2} show the relative alignment of polarization and SIG. We observe the alignment  decreases with the increase of $M_s$ which is also supported by the lower panels of Figure \ref{fig:2} where the distribution of the SIGs about the magnetic field direction is shown. We note that there exist different ways of measuring the turbulence sonic Mach number and the studies like those illustrated in Figure \ref{fig:2} allow to evaluate the accuracy of magnetic field tracing using the SIGs.

 \begin{figure*}[tbhp]
\centering
\includegraphics[width=0.98\textwidth]{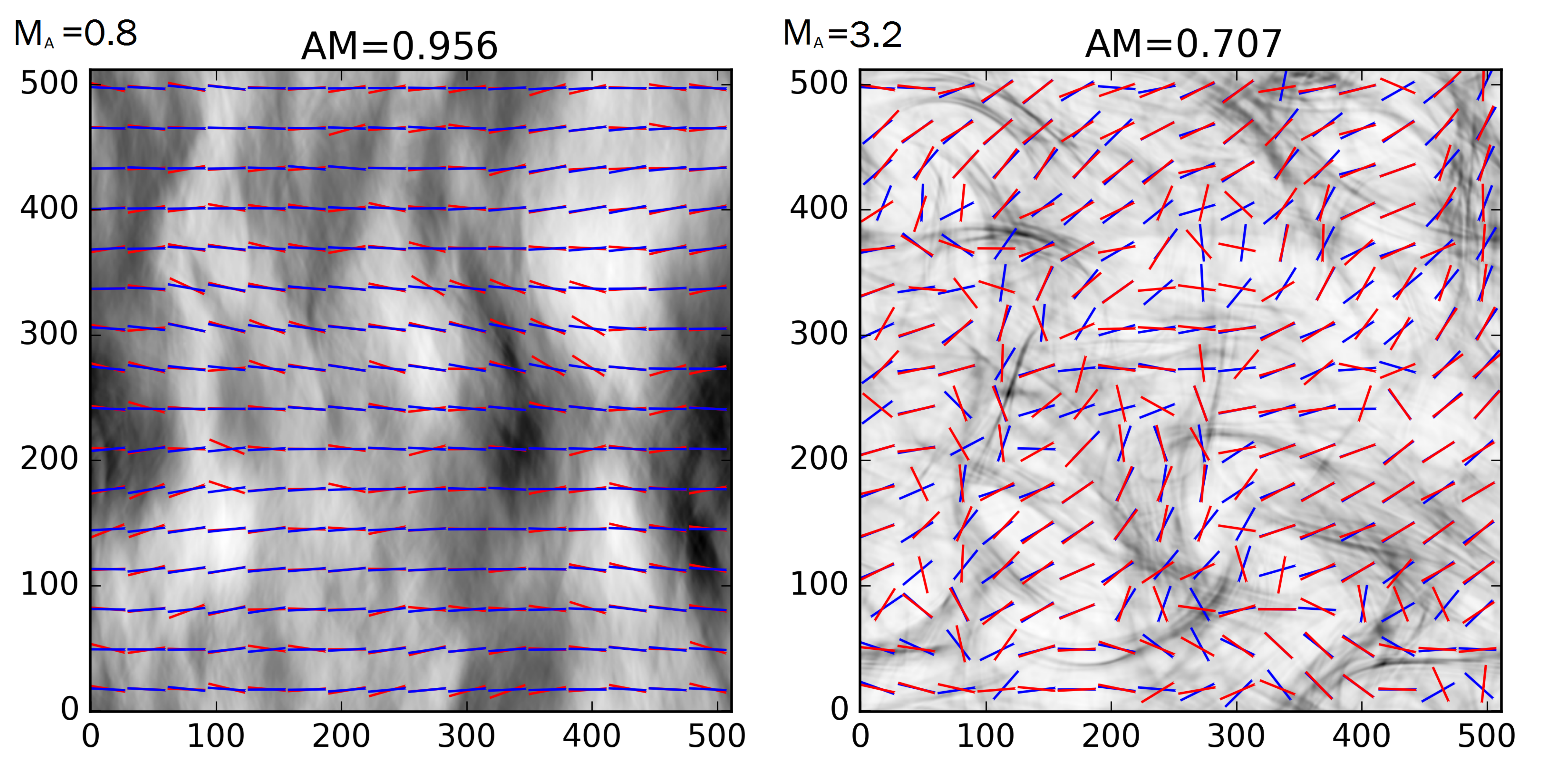}
\caption{\label{fig:superalf} SIGs(Red) and synchrotron polarization direction(Blue) in a sub-Alfvenic, incompressible (left) and a super-Alfvenic, incompressible (right) simulations, the background is the synchrotron intensity.  }
\end{figure*}

We also test the effect of Alfvenic Mach number to the alignment of SIGs on magnetic field in Figure \ref{fig:superalf}. Even with high Alfvenic Mach number ($M_A=3.2$) our method is still very good on tracing magnetic field, with $AM\sim0.71$. Not to mention, the sub-Alfven case has extremely good alignment. Notice that the alignment of SIGs in incompressible cases are significantly higher than that of the compressible cases. The possible reasons for this are e.g. creation of misaligned fast modes (See Figure \ref{fig:0}) or shocks formed due to compression of fluids.

\subsection{SIGs: effects of Gaussian noise}
\label{subsec:gf}
\begin{figure}[t]
\centering
\includegraphics[width=0.48\textwidth]{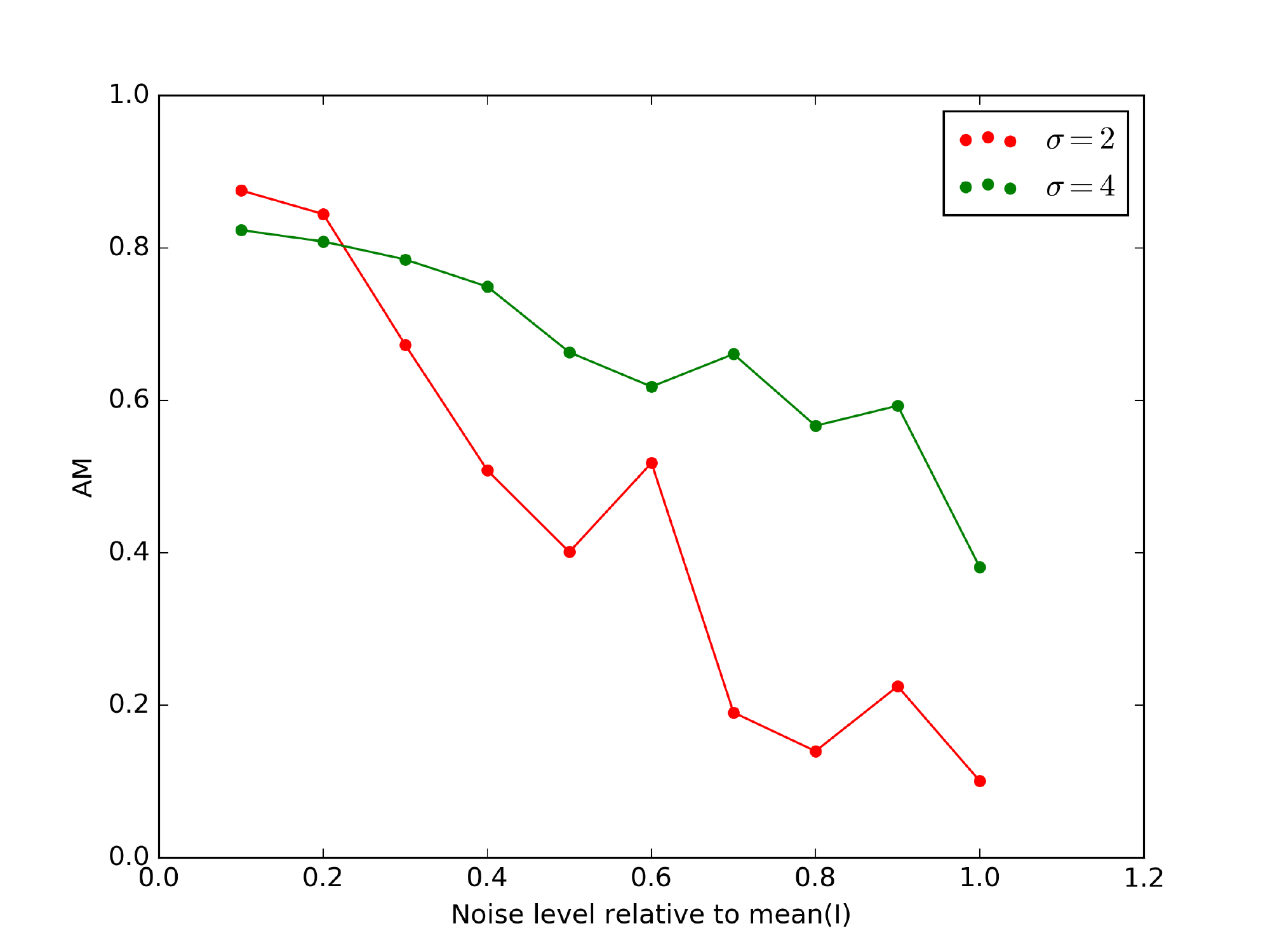}
\caption{\label{fig:3} The alignment measure of the SIGs as a function of the noise level for maps produced for $M_s=0.5$ and $M_A=0.1$ using block-size of 64, which corresponds to the middle panels of Figure \ref{fig:1}. $\sigma$ indicates the width of the Gaussian Filters, in the unit of pixels, applied in the pre-processing step.  }
\end{figure}

Real observational data is affected by noise. To gauge the effect, we test to what extend the alignment persists in the presence of noise. We calculate the SIGs by adding white noise to our synthetic maps,
The noise is included in the data in the following way. We generate white noise such that the noise amplitude is Gaussian with mean value zero. The noise level is defined as the standard deviation of the noise distribution. The resultant noise is added to the original map. The noise level is selected to be the multiple of $0.1$ of the mean synchrotron intensity, extended to a maximum equals to the mean synchrotron intensity.  

We treat the synthetic data as it if it were the real observational data. For this purpose, we analyze our noisy data using pre-processing Gaussian filters \citep{NA08}, which is a procedure frequently used as a noise reduction tool in observations. The smoothing effect from the Gaussian filter enables us to compute per-pixel gradient information more accurately. The strength of the filter is controlled by the width $\sigma$, which characterizes how many pixels are averaged to give the information of one pixel in the filtered map. A larger $\sigma$ will suppress the noise and produce a smoother map at the expense of losing the structure of magnetic field at small scales. To see the effect of the filter on the alignment, we perform tests with several $\sigma$ with the maps having different noise levels, and measure the $AM$ of the resulting maps.

The alignment measures for various noise levels and several Gaussian filters are shown in Figure \ref{fig:3}. Without the Gaussian filter pre-process, the alignment is strongly reduced, in agreement with our expectations. However, by applying Gaussian filters we can significantly improve the alignment. While for small $\sigma$ the alignment decreases rapidly with the increase of the noise, a filter with larger width improves the alignment even in a strong-noise environment. This experiment demonstrates that SIGs present a robust tool that can trace magnetic fields using observational data in the presence of noise.

\subsection{SIGs: effect of missing spatial frequencies}

To increase the resolution of available data interferometers can be used. In fact, the detailed maps of galactic synchrotron radiation as well as synchrotron emission of nearby galaxies can be obtained with interferometers. Interferometers measure the spatial Fourier components of the image and changing the baseline of the interferometer one gets different spatial frequencies. For interferometric observations, the single dish measurements deliver low spatial frequencies. The single dish observations frequencies are not always available. Then it is important to understand how this can affect the accuracy of our SIG technique.

We note that the synchrotron polarization gradients were used in \cite{Gaensler2011Low-Mach-numberGradients} and one of the motivations for their use was the possibility of using gradients with the interferometric data obtained without single dish observations. Below we test how the accuracy of the SIGs in tracing magnetic fields depends on the missing spacial frequencies.

\begin{figure}[t]
\centering
\includegraphics[width=0.48\textwidth]{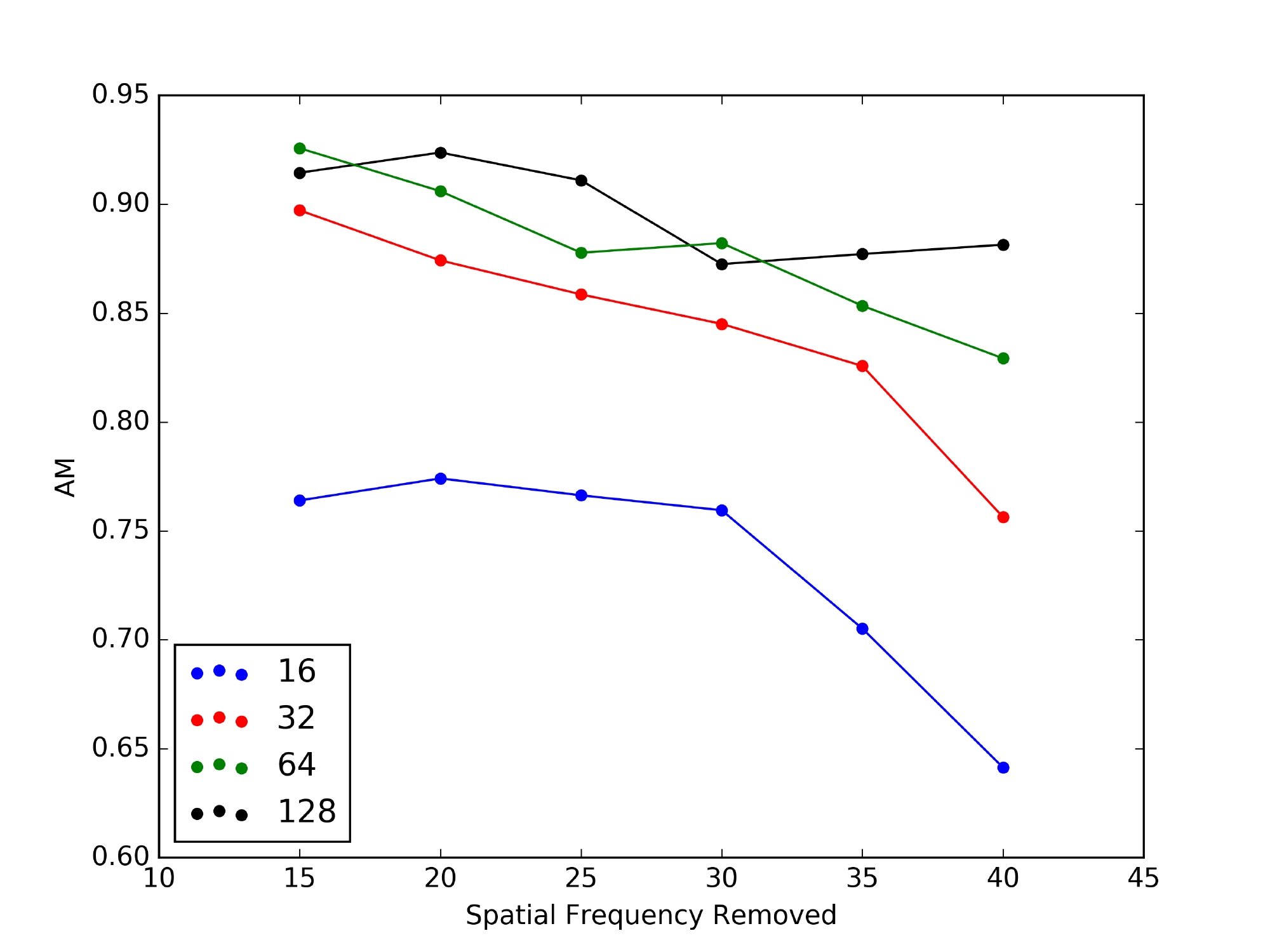}
\caption{\label{fig:4} The alignment measure of the SIGs as a function of the removed low spacial frequencies from the map with $M_s=0.5$ and $M_A=1$ using block-sizes of 16,32,64,128, and applied a Gaussian pre-filter of 4. We by default remove the injection range, which is around $k=0-9$.}
\end{figure}

In Figure \ref{fig:4} we show the alignment measure given by Eq. (\ref{AM}) using the same data as in Figure \ref{fig:3} but gradually removing spatial frequencies starting with the lowest spatial frequencies from the inertial range of our data. We observe a gradual decrease of the $AM$. We show that by increasing the block size increases we can mitigate the effect of the absence of the lower spacial frequencies in our synthetic data. 

\subsection{SIGs:effect of the amplitudes of the gradients}

\begin{figure}[tbhp]
\centering
\includegraphics[width=0.48\textwidth]{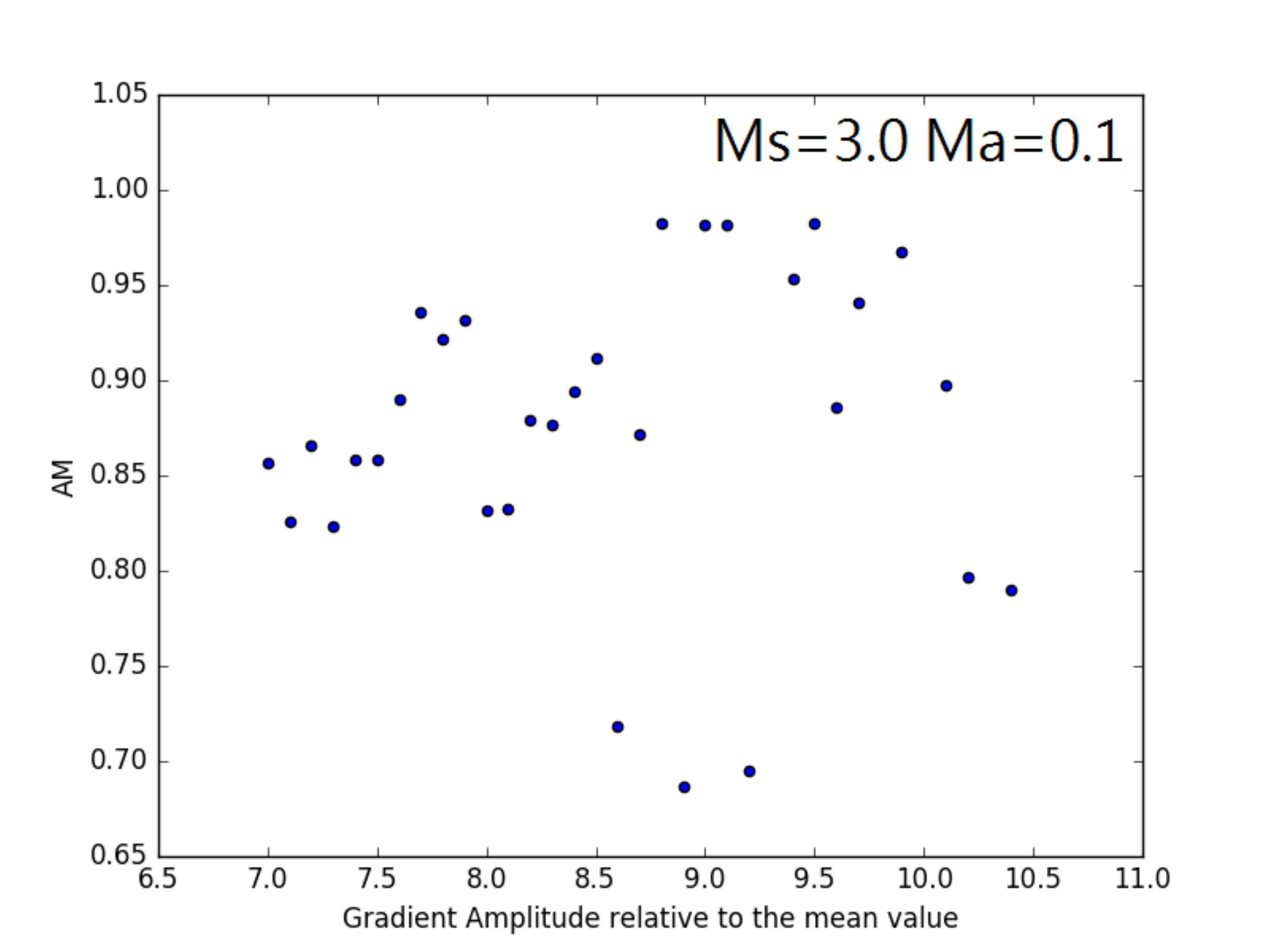}
\caption{ \label{fig:GA-AM}  A scatter plot showing the change of alignment measure relative to the change of gradient amplitude. The gradient amplitude shown here is in the multiple of mean value.  }
\end{figure}

In the present paper, similar to our earlier papers dealing with gradients (GL17, YL17, LY17), we have used the gradients to trace the magnetic field and did not account for the gradient amplitudes. 
The gradient amplitudereflects how spacial rate of the change. For example, in shock-dominated regions, we expect to see a sharp change on the magnitude across the shock boundary. On the other hand, in self-gravitated regions the gradient amplitude should increase significantly due to rapid infalling gas motions. Such events are not related to MHD turbulence and therefore one may expect that the magnetic gradients from such events are not being aligned perpendicular to the local magnetic field. At the same time, our current numerical study is limited to diffuse media, for which such a sharp changes of gradient magnitude are not common. Figure \ref{fig:GA-AM} shows how the AM changes with the gradient amplitude. With the parameters we use, we do not see a clear tendency.  The corresponding study will be done elsewhere. 

\section{Comparison with the magnetic field tracing in LP12} 
\label{sec:5}
SIGs are not the only way to trace the magnetic field with synchrotron intensity maps. For instance, anisotropic MHD turbulence also results in synchrotron anisotropies that are quantified in LP12. There the quadruple moment of the synchrotron intensity correlation functions was shown to be aligned with the magnetic field. Therefore by measuring the longer direction of the contours of the magnetic field isocorrelation (see LP12) one can approximate the magnetic-field direction over the sky.

\begin{figure*}[tbhp]
\centering
\includegraphics[width=0.48\textwidth]{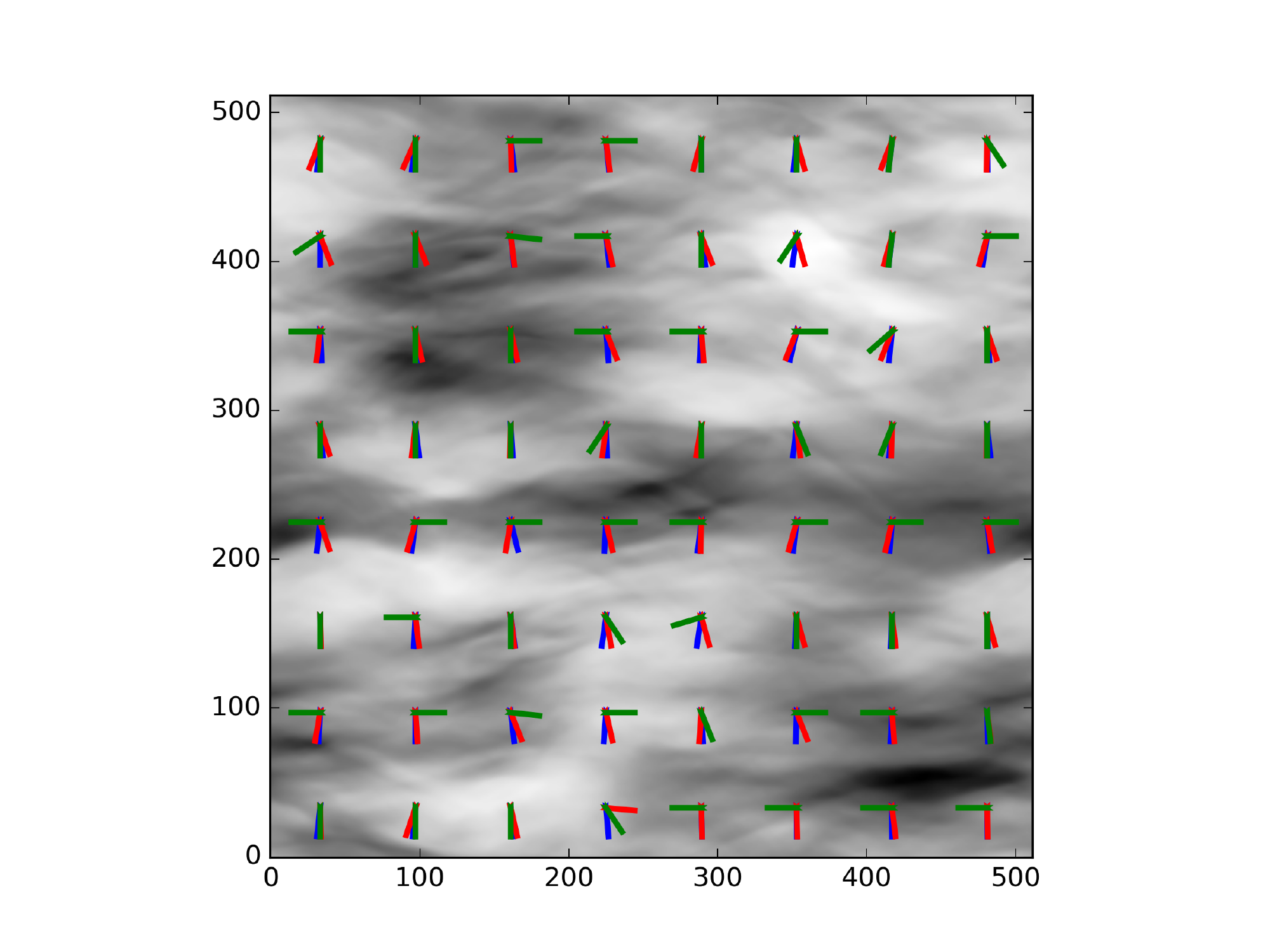}
\includegraphics[width=0.48\textwidth]{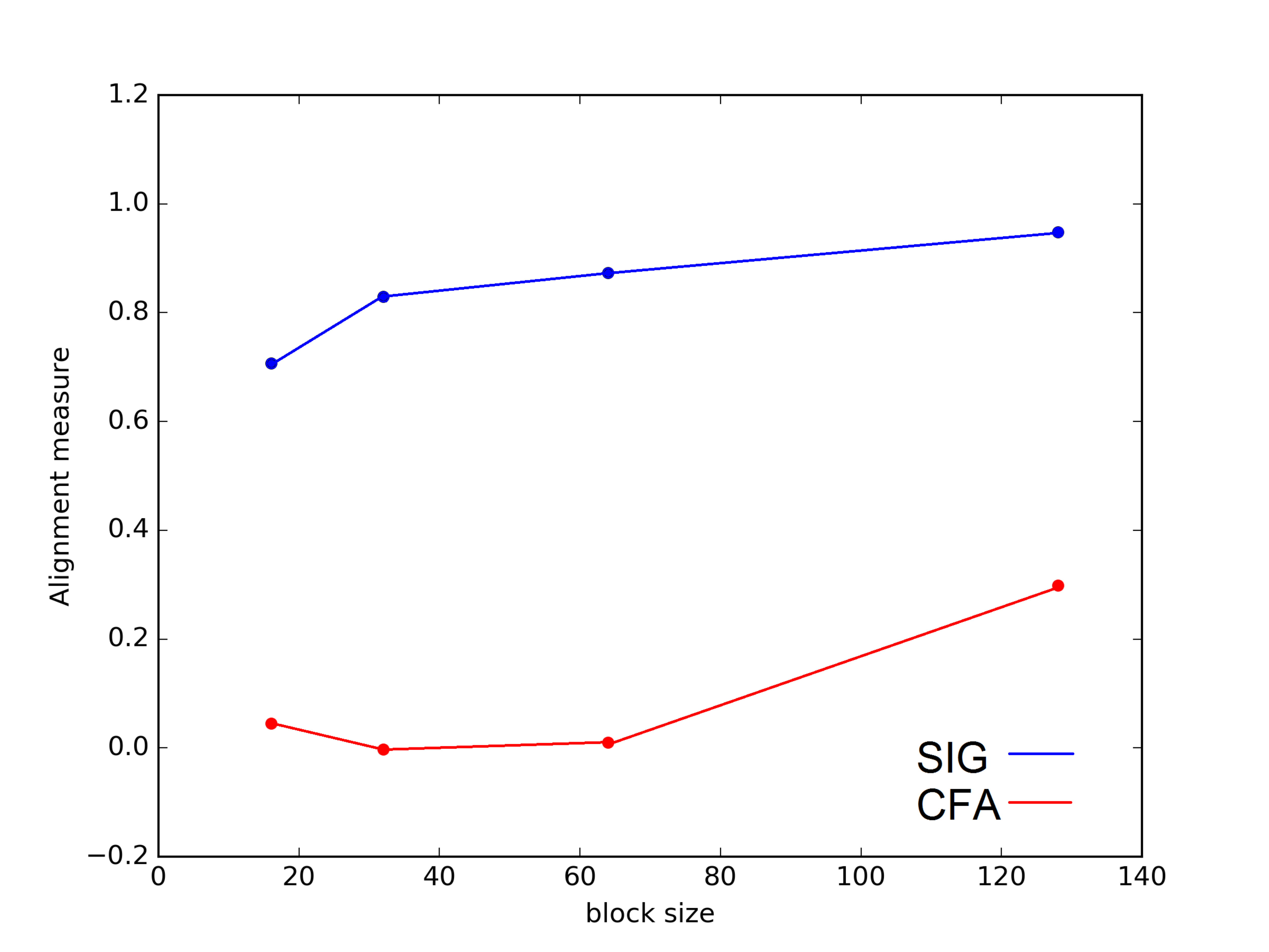}
\caption{\label{fig:6} 
Left panel: The directions of the SIGs (red) and CFA(green) are overplotted with the synchrotron polarization directions(blue) using a sub-block of size 64. Background is the synchrotron intensity, which darker color stands for higher intensity values. Right Panel: The alignment measure of the SIGs (Blue) and CFAs (Red) as a function of the block sizes after the smoothing kernel. It is obvious that SIG has advantage over the CFA for detailed tracing magnetic fields. The latter only shows a definite positive alignment measure when the block size becomes 128. }
\end{figure*}

The calculations of the correlation functions, require averaging, which in the astrophysical situations means the volume averaging. Therefore one may expect that compared to the SIGs, the LP12-type quadrupole and higher multipole anisotropies are a significantly more coarse-graded measures. 
To test this statement we provide in Figure \ref{fig:6} the $AM$ for the SIGs and the similarly-defined alignment measure of the correlation function anisotropies (CFAs) that reveal the dominant quadrupole anisotropy induced by magnetic field.  The directions of the CFAs longer axes of anisotropy are rotated 90 degrees to be compared with the directions of the SIGs and the magnetic field traced by synchrotron polarizations.

We compare sub-block averaged SIGs with the CFAs obtained in the same blocks. Figure \ref{fig:6} clearly shows that the SIGs have a great advantage over the CFAs on tracing the detailed strucuture of magnetic field. In fact, in terms of the alignment measure, the CFAs can trace magnetic field only in a for a sufficiently coarse block size. Comparatively, the SIGs can work on smaller scales without losing much of the alignment. The ability of CFAs for the same purpose is highly limited. The limitations of the CFAs compared to the SIGs is expected as the SIGs are defined for an individual eddy, while the CFAs get defined after correlation/structure functions are calculated. The latter requires the averaging over many eddies. 

We, however, believe that the SIGs and the CFAs are complementary measures in a number of ways. The correspondence between coarse-graded magnetic field directions measured by the two techniques makes the tracing of magnetic field more trustworthy. Their correspondence also indicates that the performed averaging may be sufficient to use studies of the CFA anisotropies for the purpose of separating the contribution from fundamental MHD modes, i.e. Alfven, fast and slow, as it is described in LP12.

\section{Illustration of the SIGs technique using PLANCK synchrotron data}
\label{sec:4}
The encouraging results above stimulated us to apply the SIG technique to the PLANCK synchrotron data. For our test we picked the PLANCK foreground synchrotron intensity map \citep{Planck15X} and compared the magnetic-field directions that we obtained with SIGs with the magnetic-field directions as determined by the PLANCK synchrotron polarization.

We use the full-sky map to illustrate how the SIG can trace magnetic field. We use the synchrotron intensity to compute gradients, and synchrotron polarization to infer the magnetic field direction. We projected the data into the Cartesian frame, and follow the procedures described in the earlier sections and tested with the synthetic data to produce the sub-block averaged gradient map. As shown in section \ref{subsec:gf} that $\sigma=4$ can already preserve the alignment in strong-noise environment, we reduced the noise using a $\sigma=4$ Gaussian pre-filter. 

\begin{figure*}[t]
\centering
\includegraphics[width=0.98\textwidth]{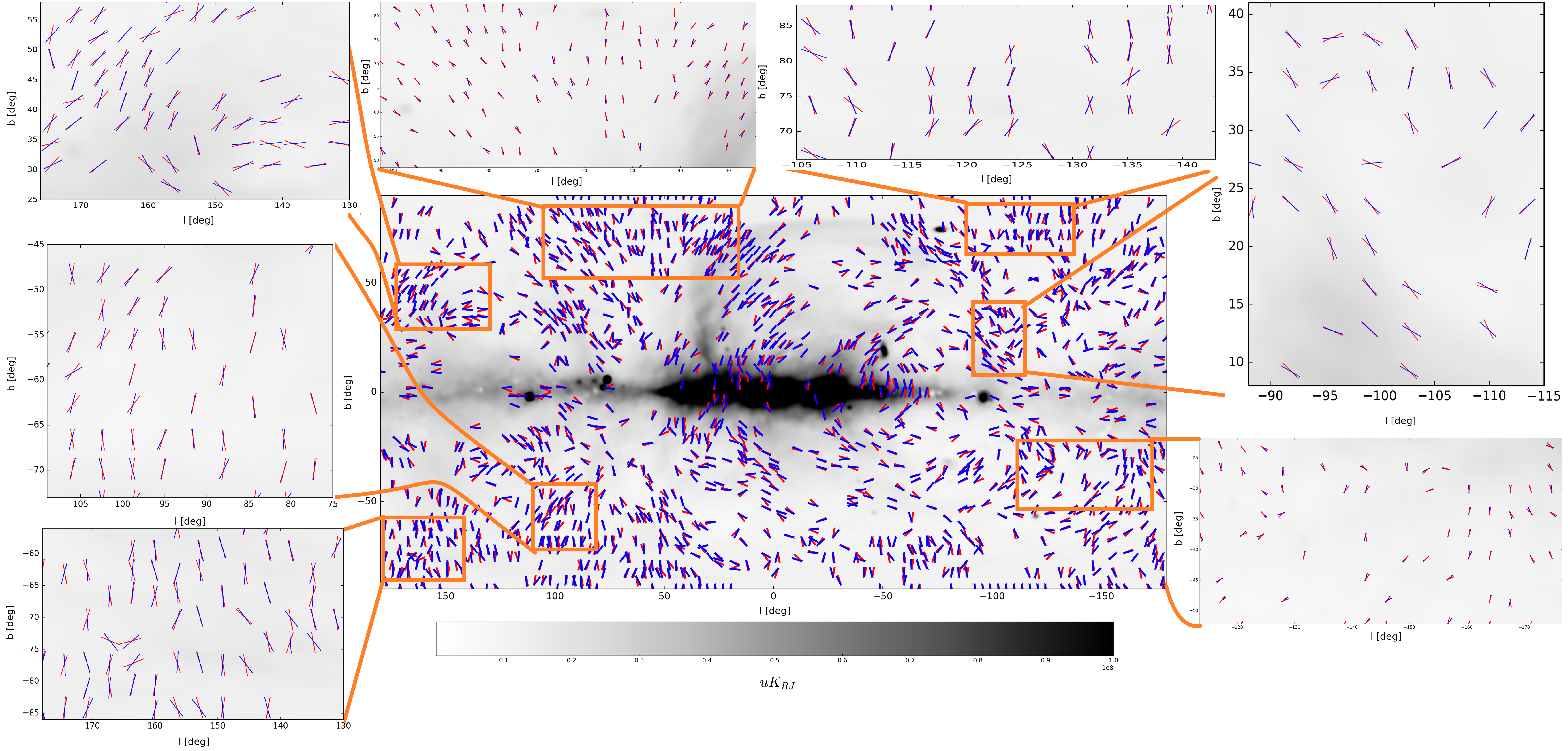}
\caption{\label{fig:5} SIGs (red) and magnetic field traced by synchrotron polarization vectors (blue) with $AM>0.5$ on the {\it full sky} Planck synchrotron data, overplotted with the synchrotron intensity when block size $= 20$, in which darker stands for higher intensity. We zoom in seven regions that are away from the galactic plane to show the relative alignment between SIGs (rotated $90^o$) and magnetic fields}
\end{figure*}

Figure \ref{fig:5} shows the full-sky SIG overplotted with Planck synchrotron data. We panels in the Figure demonstrate the patches of the sky with $AM>0.5$. It can be seen readily that the high latitude have the best alignment, while near the galactic plane alignment is reduced.  This reduction with the existence of the poorly resolved synchrotron structures not directly associated with turbulence. To such structures our technique is not applicable. However, we expect that with higher resolution when these structures are well resolved, the underlying small scale turbulence should again induce the alignment of the SIGs and the projected magnetic field.\footnote{Another potential complication related to the use of SIGs for studying magnetic field in the galactic disk plane is that distant regular structure, e.g. supernova shells, can interfere with the calculation of the SIGs that make use the resolved turbulence associated to the nearby objects. We do not discuss this effect in this paper.} At the same time, the fact that the SIGs are not influenced by the complex pattern of the Faraday rotation within the galactic disk should motivate further studies of the SIG ability to reveal magnetic fields in the galactic disk.

Our test calculations shows that the SIGs are applicable to the synchrotron intensity observations and can reveal the direction of galactic magnetic field at least at high galactic latitudes. Note, that the turbulence at such latitudes corresponds to low $M_A$.  Naturally, more tests of the SIGs in the presence of complex magnetic-field morphology are needed. Therefore moving from our demonstration here to studies of magnetic fields in the galactic disk requires a more detailed study and will be performed elsewhere.

\section{Synergy with other techniques of magnetic-field study}
\label{sec:6}
The paper above introduces a new way to trace magnetic field. It is always good to have yet another way of studying astrophysical magnetic fields. However, the advantages of the SIGs are not limited by this. 

Synchrotron polarization is a generally accepted way of studying magnetic fields in our galaxy, external galaxies and galaxy clusters. One of the difficulties of using synchrotron polarization is that the polarized radiation is subject to the Faraday rotation effect. To account for this effect, multifrequency observations are performed and the Faraday rotation is compensated. This is a significant complication. Potentially going to very high frequencies makes the Faraday effect negligible. However, at high frequencies the energy loses of relativistic electrons are not negligible and their spacial distribution may be different from the low energy relativistic electrons. This complicates the analysis and may be the source of an error. Moreover, recent analytical studies in \cite{LP16}  have demonstrated that the separation of the effects of the Faraday rotation in the presence of turbulent magnetic fields is far from trivial (see also \citealt{2016ApJ...831...77L,2016ApJ...825..154Z}). In this situation, the possibility of obtaining magnetic field direction using SIGs is very advantageous.

Combining the SIGs and the polarization measurements can be very synergetic. By measuring the actual direction of magnetic field using the SIGs and comparing it with the direction of polarization one can get the measure of the Faraday rotation of the media between us and the synchrotron-emitting region. In the presence of Faraday depolarization the combination of SIGs and polarized radiation presents additional advantages. For instance, the SIGs of the unpolarized synchrotron can trace magnetic fields in the distant regions subject to the Faraday depolarization, while the polarization and the SIGs measured for {\it polarized intensities} can trace the magnetic field in the regions close to the observer. As the Faraday depolarization is the function of frequency, by changing the frequency one can provide the 3D tomographic studies of the magnetic field structure. The corresponding procedures will be elaborated elsewhere. In particular, we are preparing a study revealing how the polarized intensity gradients trace magnetic fields.
 
The SIG technique is similar to the Velocity Centroid Gradient (VCG) technique that was introduced in GL17 and the Velocity Channel Gradient (VChG) technique introduced in LY17. Within the VCG technique, the calculation of gradients is performed using the 2D maps of velocity centroids gradients, while for the VChG technique the gradients are calculated by the intensities within the channel maps. Both measures are aimed at getting the velocity information. Indeed, the velocity centroids (see \citealt{LE03, EL05, K17}) are known to be good measures of velocity, in particular, for subsonic turbulence. Intensities in thin velocity channels are mostly dominated by velocity caustics and therefore are also most sensitive to velocities \citep{LP00}.\footnote{Naturally, this is the case when the turbulent broadening of the spectral lines is larger than the thermal broadening. Otherwise, the intensity variations arising from the velocity crowding induced by turbulence are exponentially suppressed \citep{LP04} and the channel maps reflect the total intensities.} Supersonic velocity turbulence can be studied for both HI and heavier species using velocity channel maps, while for heavier species both subsonic and supersonic velocity turbulence can be studied. Both VCGs and VChGs are readily available from the Doppler-shifted spectroscopic data. 

Compared to the VCGs and VChGs, the calculation of the SIGs is simpler, as it requires only synchrotron intensities, rather than full spectroscopic data. In this sense, the SIG technique is similar to tracing magnetic fields using intensity gradients (IGs) that are discussed as ways of tracing magnetic field and the ISM processes in GL17, YL17ab and LY17. The alignment of intensity gradients and densities was first discussed by \citep{Soler2013} on the basis of empirical numerical studies, while the relation of this alignment with the properties of MHD turbulence was revealed in the aforementioned papers. We note,  however,  that from both theoretical considerations and numerical simulations of turbulence we expect the properties of MHD turbulence to be better represented by fluctuations of magnetic fields and velocities rather than fluctuations of density (see Brandenburg \& Lazarian 2013 for a review). In particular, we expect the IGs to be more affected by shocks and be less reliable traces of magnetic field for supersonic turbulence. At the same time, the disadvantages of the IGs in tracing magnetic field present advantages in tracing other ISM processes, e.g. shocks. Thus it is really advantageous to search the synergy of different techniques, including the IGs. An apparent advantage of the IGs is that they can be used for the data sets where no spectral information is available, but only intensities. Therefore the IGs can be used e.g. with dust emission intensities. The VChGs smoothly transfer to the IG technique when the Doppler shift of the lines get subsonic or, for the sake of reducing the noise, the channels are made thicker.  

It is clear that, in general, SIGs, VCGs, VChGs and IGs are complimentary techniques that trace magnetic field in different interstellar environments. For instance, cold and warm diffuse HI, line emission, e.g. CO emission, from molecular clouds present the natural environments for studies using the VCG and VChG techniques. Combining that with the IGs, one can study shocks and self-gravitating regions (YL17, \citealt{YL17b}) and measuring the relative alignment of the directions defined by the VCGs and VChGs with those defined by the IGs one can characterize the sonic Mach number of the media (LY17).  At the same time, synchrotron radiation in the Milky Way mostly originate at the large expanses of the galactic halo and to the data from these regions the SIG technique is intended to be applied. Obtaining magnetic field properties in different parts of the interstellar media is important not only for understanding the importance of the magnetic field these phase, but also for our understanding whether the same magnetic field connects different interstellar phases. 
 In fact, while VCGs and VChGs are useful for studying magnetic fields in Cold and Warm phases, the SIGs can study magnetic fields in Warm and Hot ISM phases (see Draine 2011 for the list of the ISM phases). The advantage of the VCGs, the VChGs and the IGs is that it is possible to combine different molecular species that are present at different densities, it is possible to study the magnetic fields and the gravitational collapse within molecular clouds. In addition, using the galactic rotation curve, it is possible to approximately map the 3D distribution of magnetic field. The 3D mapping of magnetic field is also possible combining the SIGs and gradients of  polarized intensities measured at different frequencies. The latter technique requires further research, however. 
 
The alignment of interstellar dust is a well-accepted way of tracing magnetic field. Both theoretical considerations and the observational testing (see \citealt{2007JQSRT.106..225L,2015ARA&A..53..501A} and ref. therein) indicate that
the alignment of dust is very efficient in the diffuse media where radiative torques \citep{1976Ap&SS..43..291D,1996ApJ...470..551D} are strong. The alignment can trace magnetic fields in the self-gravitating regions, but it may fail in starless molecular cloud cores. The polarization arising from grain alignment is complementary to the VCGs as it is discussed in YL17 and to the VChGs, as discussed in LY17. The dust polarimetry provides the direction of the magnetic field which through the comparison with the velocity gradients reveals the regions of the gravitational collapse. Therefore, combining the polarimetry and velocity gradients it is possible to identify the regions of molecular clouds that are subject to the gravitational infall, i.e. revealing the initial stages of star formation. At the same time, compared to dust submillimeter polarimetry, velocity gradients can provide better information where the measured magnetic fields are spatially located along the line of sight. The latter is especially valuable for studying the magnetic field in the plane of the galaxy. For instance, while the polarized dust emission present in the disk plane presents the cumulative result of emission from many clouds along the line of sight, the studies with the VChGs allow mapping magnetic fields of individual clouds. In addition, the dust disk is significantly thinner that the synchrotron halo. Therefore by comparing the results obtained  FIR polarimetry with the magnetic fields revealed by synchrotron, e.g. by the SIGs, can help understanding the distribution of magnetic field with the height. Note, that in YL17 the VCGs were shown to reveal well the structure of magnetic field in the HI disk.Combining SIGs, VCGs, VChGs and the dust polarimetry one can study how magnetic fields connect Hot, Warm and Cold ISM phases with molecular clouds. 
 
In addition, we shall mention the empirical technique of tracing magnetic field using filaments observed in HI velocity channel maps \cite{2015PhRvL.115x1302C}. The relation of the observed filaments to the MHD turbulence theory requires further studies, but here we can provide some preliminary considerations. For instance, as we discussed earlier, the structures in channel maps are mostly induced by velocities and therefore we believe that the filament technique is related to velocity gradients, in particular to the VChGs.  The filaments are expected to be created perpendicular to the velocity gradients, i.e. therefore the filaments are expected to be aligned along magnetic field, which corresponds to observations. The comparison of the magnetic field tracing using the VCGs, VChGs and the filaments will be provided elsewhere. 
 
As it is clear from the discussion above, combining different velocity and synchrotron gradients, one can investigate the relative distribution of magnetic fields in different ISM phases along the line of sight. Such studies are essential for understanding of the complex dynamics of magnetized multiphase ISM. At the same time, for some regions, however, e.g. for the supernovae shocks, it seems possible and very advantageous to apply all these techniques at once.
 
 One also should note that the tracing of magnetic field directions using velocity and magnetic field gradients is different in the regions dominated by self-gravity. It was noted in \cite{YL17b} that the velocity gradients in such regions get parallel to magnetic field, as the matter falls into gravitational wells. On the contrary, we may expect the magnetic field gradients to stay perpendicular to the magnetic field even in those situations. 
 
  In addition, it is clear from the discussion in \S \ref{sec:2} that the most reliable magnetic-field tracing is expected in nearly incompressible turbulence in the absence of self-gravity. These are the conditions for the Warm and Hot phases of the ISM (see \citealt{Draine2011PhysicsMedium} for the list of the idealized ISM phases). These are exactly the media that are responsible for the bulk of  synchrotron radiation \citep{Beck15}. 
 In fact, earlier studies (e.g. \citealt{2008ApJ...686..363H, Gaensler2011Low-Mach-numberGradients}) indicated that the sonic Mach number of the synchrotron emitting Warm media is around unity. It is expected to be much less than unity for the hot coronal gas of the galactic halo. Therefore we expect that the SIGs can trace magnetic fields well and be less affected by the distortion that arises from compressibility effects. Compared to the VCGs the SIGs can be also more robust, as the VCGs are influenced by the density distribution of the emitting gas (see \citealt{EL05}) and the density is not a robust tracer of MHD turbulence statistics. VChGs when thin slices are used are marginally influenced by the turbulent densities for subsonic turbulence,\footnote{For subsonic turbulence one should use the heavier species, e.g. metals or complex molecules, that are moved by main hydrogen dominated subsonic flow. E.g. CO molecules can be used as such a tracer.} but still are affected by density for supersonic turbulence (LP00). At the same time, the synchrotron intensity fluctuations are produced by uniformly distributed electrons and thus are expected to better reflect the magnetic-field statistics. 

We can add that the ways of studying VCGs, VChGs and SIGs are similar. For instance, within our present study we successfully used the way of calculating gradients first suggested in YL17. In addition, our present study also shows that SIGs similar to VCGs and VChGs can be obtained using interferometric data with missing low spatial frequencies, e.g. the interferometric data obtained without the corresponding single dish observations. This opens prospects of using these techniques for studying extragalactic magnetic fields.

% The paragraph for fig 8.
%We also investigate how a change of the fluid compressibility $\beta=2 M_A^2/M_s^2$ affects the AM of SIG. The result is shown in Figure \ref{fig:beta-AM}. By using the cube from the ZEUS-family and the incompressible cube (See Table \ref{tab:simulationparameters}). There is a clear trend that the alignment between SIG and magnetic field direction increases as $\beta$ increases. The trend clearly illustrates compressibility of fluid will dramatically reduce the ability of SIG in tracing magnetic field. {\color{blue} While this result clearly show a trend on $\beta$ to AM, notice that it is not necessary for small $\beta$ to have very bad alignment. For example, the simulations we used from \cite{Cho2002CompressiblePlasmasb} has a significantly smaller mixing of the three modes, resulting a drastically smaller fast mode to Alfven and slow mode ratio than the cubes from the ZEUS family. As discussed before, fast modes contributes negatively in terms of the AM (see Figure \ref{fig:0}). This explains why the cubes from \cite{Cho2002CompressiblePlasmasb} has a better AM even with very small $\beta$.}

%\begin{figure}[tbhp]
%\centering
%\includegraphics[width=0.48\textwidth]{Fig8.pdf}
%\caption{\label{fig:beta-AM} The dependence of how plasma $\beta=2 M_A^2/M_s^2$ affects the alignment measure of SIG. We here include an incompressible simulation corresponding to $\beta=\infty$ to illustrate the trend. We only used the cubes from the ZEUS family and the incompressible cube for this figure. }
%\end{figure}

Faraday rotation is an important way of studying the magnetic field component parallel to the line-of-sight (see \citealt{Beck15}). The observationally attainable rotation measure (RM) is proportional to the integral of the product of the parallel to the line-of-sight component of magnetic field and thermal electron density, if the original magnetic-field direction at the source is known. The SIGs can be used to define this direction, which has advantages over the currently-used Faraday-rotation measurements that employ multifrequency polarization measurements. Moreover, SIGs can help to distinguish the Faraday rotation that arises from the source of polarized radiation and the media intervening between the source and the observer. Indeed, at the source the SIGs are measuring the actual magnetic-field direction.

%{\bf  Our work here is limited to sub- and trans-Alfvenic magnetic field. In the case when the magnetic field has different orientations at different distances along the line of sight, the procedure that we suggested can still be true. Synchrotron intensity accumulates all the contributions along the line of sight, the turbulent eddies direction contribution along the line of sight will also be stacked up accordingly. In fact, As the SIG technique relies on the validity of anisotropic eddies tracing local magnetic field directions, our prediction on the magnetic field direction is reflecting the line-of-sight-weighted magnetic field direction. This is supported by our numerical simulations where we see the variations of the magnetic field along the line of sight.
%We expect the same good correspondence between the SIGs and projected magnetic field for super-Alfvenic turbulence. Indeed, our theoretical considerations in \S \ref{sec:2} explicitly suggest that the gradients at the smallest resolved scales where turbulent motions are already subAlfvenic should dominate the measured gradients. Demonstrating this numerically is more challenging, however, as with the limited resolution of numerics it is difficult to have any inertial range for MHD-type motions that start at the scale $l_A$ and continue for smaller scales. As we feel that for the SIGs the case of superAlfvenic turbulence studies is not so essential. Therefore do not present the numerical demonstration of the abilities of the techniques for tracing magnetic field in superAlfvenic turbulence.}

A promising possibility is presented with tracing of magnetic field using aligned atoms or ions (\citealt{YL12} and ref. therein). This alignment happens for atoms/ions with fine or hyperfine structure and is induced by radiation. The Larmor precession realigns atoms/ions and thus the resulting polarization becomes dependent on the magnetic field direction. This type of alignment can potentially trace extremely weak fields in the diffuse rarefied media and we expect that this can be complementary to the SIG technique. For instance, the resulting polarization of HI arising from the atomic alignment can reveal magnetic fields. The domain of the atomic alignment are the regions of low matter density but high radiation intensity. The anisotropic radiations pumps and aligns atoms/ions, while collisions randomize spin directions.

While our discussion above has focused on the astrophysical prospects of the SIGs, the possibility of magnetic field tracing has important consequences for the CMB work. Indeed, separating of polarized foregrounds and the CMB polarization is absolutely essential for detecting and studying the enigmatic cosmological B-modes. Obtaining the actual direction of the magnetic field using the SIG technique looks very advantageous in this context, as well as combining the SIG, the VCG and the VChG measurements to weed out the polarized foreground contributions from different interstellar medium components.

\section{Summary}
\label{sec:7}
Using the theory of MHD turbulence we predicted that in magnetized flows the synchrotron intensity gradients (SIGs) are expected to reveal the magnetic field.
We successfully tested this prediction using synthetic synchrotron maps obtained with the 3D MHD compressible and incompressible simulations
as well as PLANCK synchrotron intensity and polarization data. 
The new technique is complementary to the
other ways of tracing magnetic field, which includes the traditional techniques of using synchrotron and dust polarization as well new techniques that employ velocity centroid gradients (VCGs) and velocity channel gradients (VChGs). The SIGs are giving the true direction of the magnetic field in the synchrotron-emitting volume that is not distorted by the Faraday rotation effect. Therefore,
combining the SIGs with synchrotron polarimetry measurements one can determine the Faraday rotation measure. This is useful for studying line-of-sight component of magnetic field. We have demonstrated that the SIGs are a robust measure in the presence of Gaussian noise and can be obtained with interferometeric data that is obtained without single-dish telescope observations. 

{\bf Acknowledgements.}   AL acknowledges the support of the NSF grant AST 1212096, NASA grant NNX14AJ53G as well as a distinguished visitor PVE/CAPES appointment at the Physics Graduate Program of the Federal University of Rio Grande do Norte, the INCT INEspao and Physics Graduate Program/UFRN. The stay of KHY at UW-Madison is supported by the Fulbright-Lee Fellowship. HL is supported by the research fellowship at Department of Physics, Chugnam University, Korea.

\end{document}